\documentclass[amsmath,amssymb,aps,prl,showpacs,reprint,groupedaddress]{revtex4-1}

\usepackage{graphicx}

\usepackage{color}
\usepackage{ulem}
\usepackage{natbib}

\begin{document}


\title{Possible quantum liquid crystal phases of helium monolayers}


\author{S. Nakamura$^1$}
\author{K. Matsui$^2$}
\author{T. Matsui$^2$}
\author{Hiroshi Fukuyama$^{1, 2}$}
\email[]{hiroshi@phys.s.u-tokyo.ac.jp}
\affiliation{$^{1}$Cryogenic Research Center, The University of Tokyo,\\ 2-11-16 Yayoi, Bunkyo-ku, Tokyo 113-0032, Japan}
\affiliation{$^{2}$Department of Physics, The University of Tokyo,\\ 7-3-1 Hongo, Bunkyo-ku, Tokyo 113-0033, Japan}


\date{November 4, 2016}

\begin{abstract}
The second-layer phase diagrams of $^4$He and $^3$He adsorbed on graphite are investigated.
Intrinsically rounded specific-heat anomalies are observed at 1.4 and 0.9 K, respectively, over extended density regions in between the liquid and incommensurate solid phases.
They are identified to anomalies associated with the Kosterlitz-Thouless-Halperin-Nelson-Young type two-dimensional melting. 
The prospected low temperature phase (C2 phase) is a commensurate phase or a $\textit{quantum~hexatic}$~phase with quasi-bond-orientational order, both containing $\textit{zero}$-$\textit{point}$ defectons.
In either case, this would be the first atomic realization of the $\textit{quantum~liquid~crystal}$, a new state of matter.
From the large enhancement of the melting temperature over $^3$He, we propose to assign the observed anomaly of $^4$He\,-C2 phase at 1.4 K to the hypothetical supersolid or superhexatic transition.

\end{abstract}


\maketitle


Quantum~liquid~crystal (QLC) is a novel state of matter in nature.
It is a quantum phase with partially broken rotational and/or translational symmetries and fluidity (or superfluidity) even at $T =$ 0.
The electronic nematic phase, which is conceptually one of the QLCs, is recently being studied in a variety of materials~\cite{Fradkin2010ARCMP,Frey1994PRB}.
Atomic or molecular QLCs are more intuitive and direct quantum counterparts of classical liquid~crystal. 
In bosonic systems, superfluid QLC or supersolidity is also expected.
Recently the latter possibility has intensively been explored in bulk solid $^4$He~\cite{Andreev1969JETP,Kim2004Nature,*Kim2012PRL,Balibar2008JP,*Vekhov2014PRL,*cSScomment}, and the former is proposed in cold dipolar molecule gases in two dimensions (2D)~\cite{Wu2016SciRep}.
The predicted stripe phase of superfluid $^3$He in slab geometry is a candidate for fermionic QLC~\cite{Vorontsov2007PRL,Levitin2013Science}.
However, all previous experimental attempts to detect atomic or molecular QLCs have not yet been successful or still are under debate.

Atomic monolayer of helium (He) adsorbed on a strongly attractive graphite surface provides a unique arena to investigate novel quantum phenomena of bosons ($^4$He: spinless) and fermions ($^3$He: nuclear spin $1/2$) in 2D.
Particularly the prospected commensurate phase in the second layer of He (hereafter the C2 phase) is a hopeful candidate for atomic QLC because of a delicate balance among the kinetic ($\sim$10 K), He-He interaction ($\sim$10 K) and corrugated potential energies ($\leq$ 3 K).
The commensurability here is with respect to the triangular lattice of the compressed first He layer. 
The fermionic QLC might be a new perspective for the gapless quantum spin-liquid nature~\cite{Ishida1997PRL,*Fukuyama2000PhysicaB,*Fukuyama2008JPSJ} and the anomalous thermodynamic behavior below 100 mK of the $^3$He\,-C2 phase~\cite{Matsumoto2005JLTP,*Murakawa2006AIP} alternative to the multiple spin exchange model~\cite{RogerPRL1998} or others~\cite{Fuseya2009JPSJ,*Watanabe2009JPSJ}.

If the $^4$He\,-C2 phase is a QLC, the supersolid ground state can be expected.
Eventually, two previous torsional oscillator experiments on this system observed a reentrant superfluid response as a function of density below 0.4~K~\cite{Crowell1996PRB,Shibayama2009JPhys}.
However, the identification of the phenomenon is left controversial, since the detected superfluid fractions are limited to 0.01--0.02, and the density regions where the superfluid responses are detected are not quantitatively consistent with each other. These are presumably due to the poor connectivity of microcrystallites (platelets) and the existence of heterogeneous surfaces (about 10\% of the total~\cite{Sato2012PRL}), which results in ambiguity of density scale, in Grafoil substrate they used.

So far, the $^4$He\,-C2 phase has been believed to exist from the large specific-heat anomaly observed at $T \approx 1.5$~K~\cite{Greywall1993PRB} in a narrow density range between the liquid (L2) and incommensurate solid (IC2) phases.
Although a similar anomaly has been found in $^3$He, too, at $T \approx 0.9$~K at one density~\cite{VanSciverVilches1978PRB}, other details are not known. 
Instead, the existence of the $^3$He\,-C2 phase has been accepted from the various nuclear magnetic properties at low mK~\cite{RogerPRL1998,Ishida1997PRL,*Fukuyama2000PhysicaB,*Fukuyama2008JPSJ}.

The previous belief on the $^4$He\,-C2 phase has recently been thrown into doubt by the path integral Monte Carlo (PIMC) calculation by Corboz $\textit{et~al}.$~\cite{Corboz2008PRB}.
Unlike the previous PIMC calculations~\cite{Abraham1990EPL,*Manousakis1999PRB,*Takagi2009JPhys}, they claimed the instability of the C2 phase against the L2 and IC2 phases, if zero-point (ZP) vibrations of the first layer atoms are explicitly taken into account.
Their claim raised serious questions:
Are the specific-heat peaks observed in $^4$He and $^3$He the same phenomenon related to 2D melting? 
Isn't the observed C2 phase stabilized artificially by finite size effects due to the platelet structure of exfoliated graphite substrate?

In this paper, we report results of new high-precision heat-capacity measurements of the second layers of pure $^4$He and $^3$He films at temperatures from 0.1 to 1.9~K using a \textit{ZYX} exfoliated graphite substrate.
\textit{ZYX} is known to have ten times larger platelet size (100--300~nm)~\cite{Birgeneau1982PhysicaBC,*Niimi2006PRB} than Grafoil, a substrate used in all previous works.
An average number of He atoms adsorbed on an atomically flat platelet in \textit{ZYX} is $10^6$, which is more than $10^3$ times larger than that in the simulation cell of Ref.~\citenum{Corboz2008PRB}.
We obtained unambiguous thermodynamic evidence for the existence of a distinct phase (C2 phase) between the L2 and IC2 phases regardless of system size and for that the phase exists over an extended density range in both isotopes.
Our data are consistent with the hypothesis that the C2 phase is an atomic QLC containing ZP defectons.
We also discuss the possibility that the $^4$He\,-C2 anomaly is associated with a 2D melting transition of Kosterlitz-Thouless-Halperin-Nelson-Young (KTHNY) type~\cite{Strandburg1988RMP} intertwined with superfluidity.

The experimental setup used here has been described in detail elsewhere~\cite{lt26nakamura}.
The heat capacity was measured by the heat pulse method with variable constant heat flows.
In the following we show only the heat capacity of adsorbed He films after subtracting the addendum (empty cell) and the desorption contribution (see below).
The surface area of the \textit{ZYX} substrate is $30.5\pm 0.2$~m$^2$.
The vapor pressure of sample is monitored with an {\it in situ} capacitive strain gauge.

The much larger platelet size of \textit{ZYX} than Grafoil is well demonstrated by a two times higher specific heat peak at the order-disorder transition ($T = 2.9$~K) for the $\sqrt{3}\times\sqrt{3}$ commensurate phase (C1 phase) of $^4$He adsorbed directly on graphite (see Fig. 1 of Ref.~\citenum{qfs2012nakamura,*dExperimentalComment}) and for that of $^3$He as well~\cite{Nakamura}.
The commensurability of the C1 phase is with respect to the graphite honeycomb lattice.
The critical $T$ region is also wider in \textit{ZYX} being consistent with the finite size scaling.
Despite the larger platelet size, ten times smaller specific surface area ($2$~m$^2/$g) of \textit{ZYX} causes much larger desorption heat-capacity contribution. 
This prevents us from analyzing present experimental data with reasonable accuracies at temperatures higher than $1.8$--$1.9$ and $1.3$--$1.4$~K for $^4$He and $^3$He, respectively.

Let us first show $T$ dependencies of measured heat capacities ($C$) of $^4$He films.
The data taken at densities of 17.50 $\leq \rho \leq$ 19.73~nm$^{-2}$ are shown in Fig.~\ref{fig1}(a), and those at 19.73 $\leq \rho \leq$ 21.01~nm$^{-2}$ in Fig.~\ref{fig1}(b).
Here $\rho$ is the total areal density.
Since the first layer has a much higher density or Debye temperature than the second layer, the contribution to $C$ is less than 3\% in the $T$-range we studied. 
At the lowest $\rho$ (17.50~nm$^{-2}$) the system is a uniform 2D liquid.
At high $T$, this phase is characterized by a nearly constant $C$ slightly less than the $N_2k_\text{B}$ value expected for an ideal 2D gas as well as a broad maximum near 0.9~K below which $C$ rapidly falls down~\cite{Greywall1993PRB,Nakamura}.
Here $N_2$ is the number of He atoms in the second layer which is calculated from the known first-layer density ($\rho_1$) vs $\rho$ relation obtained from the neutron scattering data~\cite{Lauter1987CanJP,*Lauter1991PTSF} using the second-layer promotion density of $11.8\pm0.3$~nm$^{-2}$~\cite{qfs2012nakamura}.
For example, $\rho_1 = 12.05\pm0.3$~nm$^{-2}$ at $\rho =$ 19.73~nm$^{-2}$.

\begin{figure}[t]
\includegraphics[width=.48\textwidth]{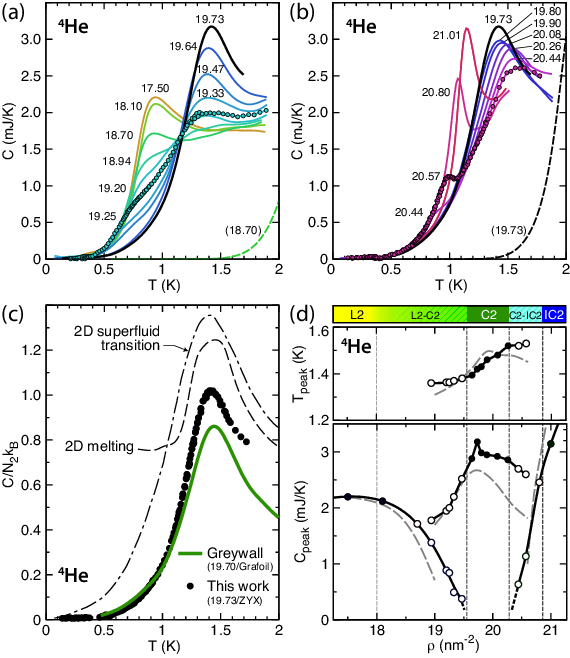}%
\caption{(a), (b) Heat capacities of the second layer of $^4$He on \textit{ZYX} graphite. 
The numbers are total densities in nm$^{-2}$. 
For clarity actual data points are plotted only for $19.25$ and $20.57$~nm$^{-2}$. 
The dashed lines are desorption contributions which have already been subtracted from the raw data. 
(c) Specific heat of the C2 phase obtained with \textit{ZYX} (filled circles: this work) and Grafoil (solid line: Ref.~\citenum{Greywall1993PRB}) substrates. 
Also shown are calculated specific heats for 2D melting (dashed line: Ref.~\citenum{WierschemManousakis2011PRB}) and superfluid transition (dash-dotted line: Ref.~\citenum{Ceperley1989PRB}), in which the $T$ scales are normalized by $T_{\mathrm{peak}}$.
(d) Density variations of $C_{\mathrm{peak}}$ (lower panel) and $T_{\mathrm{peak}}$ (upper panel).
The filled symbols are for the pure L2, C2, and IC2 phases and open ones for the coexistence regions.
The dashed lines are the data of Ref.~\citenum{Greywall1993PRB} adjusted to our surface area.
\label{fig1}}
\end{figure}

As $\rho$ increases above $18.70$~nm$^{-2}$, a new $C$ anomaly starts to develop near $T =1.4$~K, whereas the liquid component gradually decreases.
The two features coexist until $19.47$~nm$^{-2}$.
Above $19.64$~nm$^{-2}$ the liquid component completely disappears leaving only the rounded peak at  $1.4$~K which corresponds to the C2 peak observed by Greywall~\cite{Greywall1993PRB} using Grafoil substrate.
As we further increase $\rho$, the heat-capacity peak height ($C_{\mathrm{peak}}$) becomes largest at 19.73~nm$^{-2}$ ($\equiv$ $\rho_{\mathrm{C2}}$) and then begins to decrease.
In Fig.~\ref{fig1}(c) the specific heat ($c$) data at $\rho = \rho_{\mathrm{C2}}$ obtained with \textit{ZYX} and Grafoil are compared.
They look similar except that the \textit{ZYX} data give a slightly larger $c$ around the peak temperature $T_{\mathrm{peak}}$ by about $13\%$.
Above 20.44~nm$^{-2}$ a new peak appears near $0.8$~K.
With increasing $\rho$, the peak grows rapidly in height and temperature up to $1.2$~K coexisting with the C2 anomaly which diminishes gradually keeping $T_{\mathrm{peak}}$ fixed. 
The two features apparently coexist at least until $20.80$~nm$^{-2}$.
This last peak is associated with the melting transition of the IC2 solid~\cite{Greywall1993PRB,Nakamura}.

In Fig.~\ref{fig1}(d) we plot density variations of $C_{\mathrm{peak}}$ and $T_{\mathrm{peak}}$ as well as those of Greywall~\cite{Greywall1993PRB} (dashed lines) who used Grafoil substrate.
The phase diagram determined in this work is also shown on the top.
Unambiguously, there exists a distinct C2 phase over an extended density region from $19.6$ to $20.3$~nm$^{-2}$ where we observed only the C2 anomaly (closed circles).
The C2 phase is definitely not an experimental artifact caused by finite size effects of substrate since the $c$ anomaly is even enhanced slightly with increasing the platelet size by an order of magnitude.
Within this C2 region, $T_{\mathrm{peak}}$ increases by $10\%$.
The C2 phase is well separated from the L2 and IC2 phases by L2-C2 (18 $< \rho <$ 19.6~nm$^{-2}$) and C2-IC2 (20.3 $< \rho < 20.9$~nm$^{-2}$) coexistence regions where we observed the double anomaly feature (open circles).
Although the feature is vaguely visible in Greywall's data at 19.00 and 20.30~nm$^{-2}$ (see Fig.~3 of Ref.~\citenum{Greywall1993PRB}), it is much clearer with great details here thanks to finer $\rho$ and $T$ grids and the better substrate quality.
For example, when $\rho$ approaches $\rho_{\mathrm{C2}}$ from both directions, the L2 and IC2 anomalies destruct preferentially from higher-$T$ envelopes keeping common low-$T$ envelopes, while the C2 anomaly grows without changing its $T_{\mathrm{peak}}$ so much.
This unusual behavior can never be expected from the conventional phase separation or domain wall structures.

Next we show heat capacity data of the second layer of $^3$He films at densities of $17.50 \leq \rho \leq 19.00$~nm$^{-2}$ in Fig.~\ref{fig2}(a) and $19.00 \leq \rho \leq 20.40$~nm$^{-2}$ in Fig.~\ref{fig2}(b).
The density evolution is qualitatively similar to that in $^4$He.
We observed again a clear C2 peak which becomes maximum at $\rho_{\mathrm{C2}} = 19.1 \pm0.1$~nm$^{-2}$ and $T_{\mathrm{peak}} =1.0$--$1.1$~K.
This $c$ peak is very similar to that observed by Van Sciver and Vilches using Grafoil substrate~\cite{VanSciverVilches1978PRB} as compared in Fig.~\ref{fig2}(c), indicating almost no size effects.
Here we estimated $N_2 $ assuming $\rho_1 = 11.6$~nm$^{-2}$.
The $\rho_1$ value is evaluated from the second-layer promotion density ($= 11.2\pm$0.2~nm$^{-2}$)~\cite{Nakamura} and the subsequent first-layer compression by 4\%~\cite{Lauter1987CanJP,*Lauter1991PTSF}.

In Fig.~\ref{fig2}(d) we plot density variations of $C_{\mathrm{peak}}$ and $T_{\mathrm{peak}}$ for $^3$He as well as a proposed phase diagram at $T = 0$.
The determination of each phase boundary is somewhat ambiguous compared to $^4$He, because the density grid of measurement is not fine enough here.
The double anomaly feature in the coexistence regions is hardly visible due to weakly $T$-dependent large $C$ contributions from Fermi liquids in the second and third layers.
These contributions are represented by the heat capacity isotherm at $T =0.2$~K plotted in Fig.~\ref{fig2}(d).
Note that the third layer promotion in $^3$He occurs at a relatively low density ($\approx 19.3$~nm$^{-2}$) before the C2-IC2 coexistence starts~\cite{Greywall1990PRB,*eGreywallComment,Nakamura}.
This is in sharp contrast to the case of $^4$He where the third layer promotion occurs at much higher densities ($\rho \geq$ 21~nm$^{-2}$) after the C2-IC2 coexistence completes~\cite{Greywall1993PRB}.
Also plotted in this figure is the magnetization isotherm taken at $T =$ 4.6~mK with Grafoil by B\"auerle $\textit{et~al}$.~\cite{Bauerle1996Czech}.
The agreement with our phase diagram is remarkable in terms of the density width for the C2 phase and of the third-layer promotion density.
The previous workers' data with Grafoil are consistent with our phase diagram if their density scales are multiplied by $1.015$ (Refs.~\citenum{VanSciverVilches1978PRB,VanSciver1978PRB}), $1.04$ (Ref.~\citenum{Greywall1990PRB,*eGreywallComment}), and $1.02$ (Ref.~\citenum{Bauerle1996Czech}).

\begin{figure}[b]
\includegraphics[width=.48\textwidth]{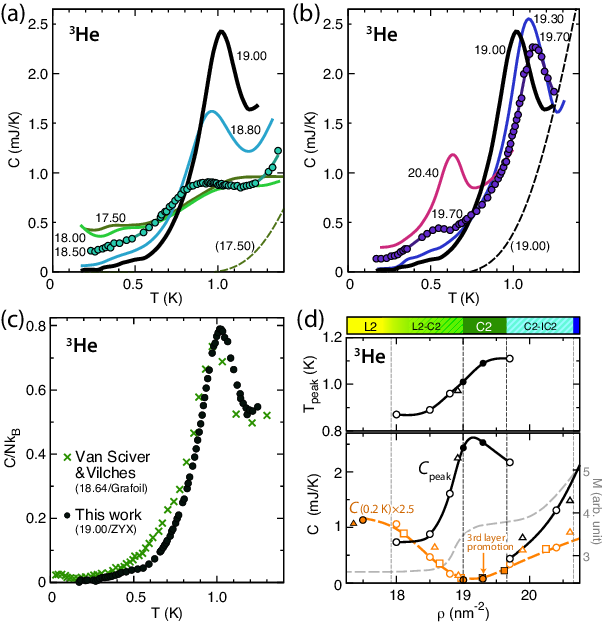}%
\caption{(a), (b) Heat capacities of the second layer of $^3$He on \textit{ZYX}. 
(c) Specific heat of the C2 phase obtained with \textit{ZYX} (filled circles: this work) and Grafoil (crosses: Ref.~\citenum{VanSciverVilches1978PRB}). 
The latter show slightly larger specific heats at low-$T$ due to a small amount of remnant L2 phase.
(d) Density variations of  $C_{\mathrm{peak}}$ (lower panel) and $T_{\mathrm{peak}}$ (upper panel).
The dashed line is a heat capacity isotherm at $T = 0.2$~K.
The triangles (Refs.~\citenum{VanSciverVilches1978PRB,VanSciver1978PRB}) and squares (Ref.~\citenum{Greywall1990PRB,*eGreywallComment}) are the data with Grafoil adjusted to our surface area and density scale (see text).
The dashed line is the magnetization at $T =$ 4.6~mK with Grafoil (Ref.~\citenum{Bauerle1996Czech}).
Other details are the same as Fig.~\ref{fig1}.
\label{fig2}}
\end{figure}

There are several reasons to believe that the broad specific-heat anomalies  observed in $^4$He and $^3$He are due to $\textit{continuous}$ 2D melting transitions of the C2 phase which possesses some type of spatial order.
For example, entropy changes associated with them, $\Delta S =$ 0.4--0.5 $N_2k_\text{B}$, are large enough to convince one of the presence of phase transitions [see Fig.~\ref{fig3}(a)].
The anomalies are substantially broad, and the broadness should be intrinsic since they are insensitive to the system size.
Eventually, they are similar to the Monte Carlo simulation for melting of a 2D classical Lennard-Jones solid~\cite{WierschemManousakis2011PRB} as shown in Fig.~\ref{fig1}(c).
This simulation is consistent with the KTHNY theory~\cite{Halperin-Nelson1978PRL,*NelsonHalperin1979PRB,*Young1979PRB}.
The theory predicts that, in general, 2D solid melts through two continuous transitions on warming, i.e., a transition from solid to hexatic at the melting temperature ($T_{\mathrm{m}}$) and that from hexatic to uniform liquid at a slightly higher temperature ($T_{\mathrm{i}}$).
The two transitions are due to unbinding of dislocation pairs and disclination ones, respectively. 
They are accompanied by intrinsically rounded specific-heat anomalies because of subsequent proliferation of the free topological defects.
The two anomalies, however, can easily merge into a broader single peak because usually $(T_{\mathrm{i}}-T_{\mathrm{m}})/T_{\mathrm{m}}\ll1$~\cite{WierschemManousakis2011PRB}.
In the same figure we also plot the PIMC simulation of the specific heat anomaly for the superfluid transition of 2D liquid $^4$He~\cite{Ceperley1989PRB} as a typical example of the Kosterlitz-Thouless type transition where a single step transition occurs due to unbinding of vortex pairs~\cite{jKentcomment}.
Importantly, these specific heat peaks are centered at 10\%--30\% higher temperatures than the true transition temperatures at which practically no anomalies are observed~\cite{WierschemManousakis2011PRB,Ceperley1989PRB,BerkerNelson1979PRB}.

Let us discuss the nature of the C2 phase in more detail.
If the periodic potential from the first layer plays an essential role, it would be a commensurate phase as was originally anticipated~\cite{bDensityRatioComment}.
If so, we should assume the existence of a sizable amount of ZP defectons such as ZP vacancies, interstitials, and dislocation pairs.
These pointlike defectons hop around quantum mechanically without breaking the translational quasi-long-range order.
Otherwise, it is difficult to explain why the commensurate phase can be stabilized over a relatively wide density region of $w \equiv \Delta \rho_{\mathrm{C2}}/\rho_{\mathrm{C2}}\approx$~0.08--0.09 where we observed only the C2 anomaly and the growth of $T_{\mathrm{peak}}$ with increasing $\rho$.
The neutron scattering data~\cite{Lauter1987CanJP,*Lauter1991PTSF} with \textit{ZYX} substrate show a small but steep increase by 1.3\%--1.5\% of $\rho_1$ at densities near the L2-C2 coexistence and the C2 regions for both $^4$He and $^3$He.
This suggests a simultaneous compression of the first and second layers supporting the commensurate phase picture.
It should, however, be noted that the neutron experiments were unsuccessful to detect Bragg peaks directly from the C2 phase, though they may be masked with a large Debye-Waller factor due to strong quantum fluctuations.
Another fundamental question of this picture is that, if the substrate potential corrugation is large enough to stabilize a commensurate phase, the transition nature should be modified so as to be size dependent acquiring Ising or first-order character, which is not consistent with our observation~\cite{Halperin-Nelson1978PRL,*NelsonHalperin1979PRB,*Young1979PRB}.

If the role of the periodic potential from underlayers is not important, the most plausible phase at least for the $^4$He\,-C2 phase is the $\textit{quantum}$ hexatic phase as was previously examined theoretically~\cite{Mullen1994PRL,Apaja2008EPL,*fHexaticComment}.
This is a quantum version of the hexatic phase where bound disclination pairs are spontaneously created by quantum fluctuations even at $T =$ 0, i.e., $T_{\mathrm{m}} \rightarrow 0$, destroying the translational order but keeping the quasi long-range sixfold bond-orientational order.
This picture is consistent with the existing experimental and theoretical constraints, e.g., the lack of translational long-range order in the neutron experiments~\cite{Lauter1987CanJP,*Lauter1991PTSF} and the recent PIMC calculation~\cite{Corboz2008PRB}.
Also, the common low-$T$ envelopes observed in the L2-C2 and C2-IC2 coexistence regions of $^4$He can be interpreted by considering that the short-range C2 (or hexatic) correlation does not affect the excitation spectrum at low momenta but softens it at a high momentum just like the $\textit{roton}$-type softening near solidification computed for 2D liquid $^4$He~\cite{Halinen2000JLTP,*KrotscheckJLTP2015}.

As far as we know, there are no other reasonable scenarios for the C2 phase other than the above-mentioned two. 
In any case, it would be a new state of matter never experimentally observed before.
It is difficult to discriminate them only from the present thermodynamic measurements, although the quantum hexatic picture looks more feasible.
New scattering experiments and first-principles calculations to measure the angular correlation function are highly desirable.
One such attempt has recently been made by Ahn $\textit{et~al}$.~\cite{Ahn2016PRB} who showed that from a new PIMC calculation the 4/7 phase is stable or unstable as a result indeed of a delicate energy balance depending on whether  the first layer is a commensurate or incommensurate solid.
In the following, we focus on differences between the data of the two isotopes which obey different quantum statistics.

\begin{figure}[b]
\includegraphics[width=.49\textwidth]{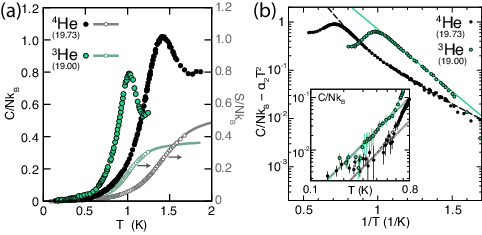}%
\caption{Specific-heat anomalies of the $^4$He- and $^3$He\,-C2 phases plotted in different fashions. (a) Linear-linear plot of the specific heat and the entropy changes deduced from them. The lines are extrapolations assuming mirror symmetry of $C (T)$ about $T = T_{\mathrm{peak}}$. (b) Inset: Log-log plot of the low-$T$ specific heat. The dotted lines are $C \propto T^2$ behaviors. Main: Arrhenius plot of the specific heat after subtracting the $T^2$ term. The solid lines are fittings to $C/N_{2}k_{\text{B}} = \alpha_2 T^2+\beta_0 \exp(-\epsilon_0/T)$ with $\alpha_2 = 0.154(4)$~K$^{-2}$, $\beta_0 = 321(33)$ and $\epsilon_0 = 6.10(8)$~K for $^3$He. The dashed lines are fittings to $C/N_{2}k_{\text{B}} = \alpha_{2}T^2 + \beta_1\exp(-\epsilon_{1}/T)+\beta_{2}\exp(-\epsilon_{2/}T)$ with $\alpha_2 = 0.058(4)$~K$^{-2}$, $\beta_1 = 1020(437)$, $\epsilon_1 = 9.85(63)$~K, $\beta_2 = 5.1(15)$, and $\epsilon_2 = 3.69(21)$~K for $^4$He.
\label{fig3}}
\end{figure}

$T_{\mathrm{peak}}$ and $\Delta S$ are larger by 40\% in $^4$He compared to $^3$He despite their nearly the same areal densities [Fig.~\ref{fig3}(a)].
This is quite singular since, in other 2D and three dimensional (3D) He solids, heavier isotopes ``always" have slightly lower $T_{\mathrm{m}}$ (or $T_{\mathrm{peak}}$) than lighter ones with the same densities~\cite{gSMComment}.
Clearly an extra degree of freedom stiffens the C2 order for the case of bosonic $^4$He, which is most likely superfluid order.
If so, it would be a $\textit{superhexatic}$ state~\cite{Mullen1994PRL,Apaja2008EPL,*fHexaticComment} (or a supercrystal with the ZP defectons~\cite{Crowell1996PRB,Shibayama2009JPhys,Andreev1982PLTP}) where the superfluidity and spatial orders coexist.
The recent topological argument on 2D QLCs~\cite{Gopalakrishnan2013PRL} indeed indicates that, by intertwining the superfluid order with the orientational one, $T_{\mathrm{i}}$ in triangular or hexagonal lattices can be enhanced by the same amount of the additional stiffness of order compared to untwined cases.
Such an enhancement does not occur in stripes and square lattices, which is another reason why we chose the hexatic order among various other spatial orders~\cite{gSMComment}.

It is expected that the specific heat of the C2 phase is dominated by a $T^2$ term which comes from 2D phonons at low $T$.
Indeed, the $^3$He data follow the $T^2$ behavior as shown in the inset of Fig.~\ref{fig3}(b).
On the other hand, it is not clear if the low $T$ $^4$He data obey the $T^2$ law because of limited accuracies due to the small surface area of the \textit{ZYX} substrate, 
Nevertheless, they are obviously much smaller (at least by a factor of 3) than $^3$He in a $T$ range between 0.3 and 0.6~K where we determined the coefficient $\alpha_2$ of the $T^2$ term.
This is again singular since, in other 2D and 3D solids of He, heavier $^4$He have slightly larger phonon terms than lighter $^3$He with the same densities~\cite{gSMComment}. 
The $^4$He\,-C2 phase seems to have much reduced spatial order than the $^3$He counterpart.
Also, the two isotopes have qualitatively different $T$ dependencies at 0.2 $\leq T/T_{\mathrm{peak}} \leq$ 0.85.
The simple Arrhenius type function (solid lines) represents the $^3$He data very well, indicating one dominant topological defect, i.e., a disclination (dislocation) pair for the hexatic (triangular) order.
However, it does not work for the $^4$He data [Fig.~\ref{fig3}(b)] suggesting involvement of another defect, in other words, another order.
These would provide useful information for future investigations of quantum fluctuations and thermal activations of the ZP defectons~\cite{gSMComment}.
For example, the activation energy is considered as twice the defect core energy.

In summary, we determined detailed phase diagrams of the second layer of $^4$He and $^3$He showing the unambiguous existence of distinct phase (C2 phase) at densities of finite span between the liquid and incommensurate solid phases.
Our data strongly suggest that the phase is either the commensurate phase or the quantum hexatic phase, which are both atomic QLCs containing the zero-point defectons. 
We also discussed the possibility of a superfluid QLC state in the $^4$He\,-C2 phase where both spatial and gauge symmetries are cooperatively broken at temperatures below 1.4~K.

\begin{acknowledgments}
We thank Tony Leggett, Hans Lauter, Luciano Reatto, Naoki Kawashima, and Henri Godfrin  for helpful discussions.
This work was financially supported by Grant-in-Aid for Scientific Research on Priority Areas (Grant No.~17071002) from MEXT, Japan and Scientific Research (A) (Grant No.~22244042) and (B) (Grant No.~15H03684), and Challenging Exploratory Research (Grant No.~15K13398) from JSPS.
S. N. acknowledges support from the Fuuju-kai Fellowship. 
\end{acknowledgments}



%

\newpage

\begin{center}
\maketitle{\bf{\large{Supplemental Material for\\
``Possible quantum liquid crystal phases of helium monolayers"}}}

S. Nakamura$^1$, K. Matsui$^2$, T. Matsui$^2$,\\
and Hiroshi Fukuyama$^{1, 2*}$\\

\small{$^{1}$\it{Cryogenic Research Center, The University of Tokyo,\\ 2-11-16 Yayoi, Bunkyo-ku, Tokyo 113-0032, Japan}}\\
\small{$^{2}$\it{Department of Physics, The University of Tokyo,\\ 7-3-1 Hongo, Bunkyo-ku, Tokyo 113-0033, Japan}}

\small{(Dated: November 4, 2016)}\\
\end{center}

\maketitle


\subsection{Isotope effect on melting of quantum solids}
The negative isotope effect on melting temperature ($T_{\mathrm{m}}$), i.e., lighter isotopes have higher $T_{\mathrm{m}}$ ($dT_{\mathrm{m}}/dm \leq 0$) is one of the most peculiar and universal properties of quantum solids.
Here $m$ is the atomic/molecular mass. 
The underlying mechanism is reduction of the quantum mechanical zero-point energy ($K$) by localization when the mean interparticle distance becomes comparable to the hardcore diameter upon compression.
This has been explained qualitatively by Nagaoka~\cite{Nagaoka1980PTPS} for hardcore particles in three dimensions (3D) at $T =$~0, numerically by Ceperley $\it{et~al}$.~\cite{Ceperley1996PRL} for hcp $^4$He at $T \neq$~0 and numerically by Hirashima $\it{et~al}$~\cite{Hirashima2003JPSJ} for quantum particles with short-ranged repulsive interactions in 2D under a periodic external potential.
Experimentally, the negative isotope effect is eventually hold in all previously known 2D and 3D solids of helium as shown in TABLE~\ref{tabI}.

Therefore, the very large (40\%) and ``positive'' isotope effect on $T_{\mathrm{peak}}$ in the C2-phase we found in this experiment is extremely anomalous. 
Statistics should play important roles on quantum melting at temperatures below about 2~K. 
In principle, superfluidity can substantially reduce $K$ in the C2 phase of bosonic $^4$He resulting in enhanced $\it{melting}$ temperature. 
On the other hand, it cannot be expected for fermionic $^3$He at this temperature range because the superfluid transition should be of the order of millikelvin or less if it exists due to Cooper pairing. 
Otherwise, we do not have any reasonable explanations for this anomalous isotope effect on $T_{\mathrm{peak}}$.

\subsection{Low temperature specific heat of the C2 phase}
In order to extract information on excitations in the C2 phase, we tried to fit the low $T$ specific heat data below $T_{\mathrm{peak}}$ to the following four functions depending on the $T$ range:
\renewcommand{\theequation}{S\arabic{equation}}
\begin{eqnarray}
\label{eq1}
C/N_{2}k_{\text{B}} &=& \alpha_2 T^2,\\
\label{eq2}
C/N_{2}k_{\text{B}} &=& \alpha_2 T^2 + \beta_0 \exp(-\epsilon_0/T),\\
\label{eq3}
C/N_{2}k_{\text{B}} &=& \alpha_2 T^2 + \beta_1 \exp(-\epsilon_1/T)  \nonumber \\
 & & + \beta_2 \exp(-\epsilon_2/T),\\
\label{eq4}
C/N_{2}k_{\text{B}} &=& \alpha_2 T^2 + \alpha_n T^n.
\end{eqnarray}

The $T^2$ terms are from 2D phonons.
Note that the lowest $T$ of our measurement is 200 mK where the magnetic specific heat associated with the nuclear spin degrees of freedom in the $^3$He-C2 phase is negligibly small~\cite{VanSciver1978PRB_SM}. 
We first fitted the data at $0.2 \leq T\leq 0.6$~K ($^3$He) and $0.3 \leq T\leq 0.5$~K ($^4$He) to Eq.~(\ref{eq1}) to determine the $\alpha_2$ values.
After substituting them into Eqs.~(\ref{eq2}--\ref{eq4}), the data in a wider $T$ range of 0.2 $\leq T/T_{\mathrm{peak}} \leq$ 0.85 were fitted with those three functions.

The $^3$He data at $\rho = 19.00$ nm$^{-2}$ are best fitted to Eq.~(\ref{eq2}) with $\alpha_2 = 0.154\pm0.004$~K$^{-2}$, $\beta_0 = 321\pm33$ and $\epsilon_0 = 6.10\pm0.08$~K. 
Generally, the simple Arrhenius type behavior such as Eq.~(\ref{eq2}) should be applicable to the low-$T$ specific heat of KTHNY transitions (or superfluid KT transitions) where unbound free topological defects are thermally excited~\cite{WierschemManousakis2011PRB_SM,BerkerNelson1979PRB_SM}.
In the quantum hexatic scenario, $\epsilon_0$ is about twice the disclination core energy.

The situation for $^4$He is rather different.
Equation~(\ref{eq3}) gives apparently better fitting than Eq.~(\ref{eq2}) indicating the existence of another type of defect with a lower binding energy than $\epsilon_0$.
This can clearly be seen in the main figure of Fig. 3(b) of the main text.
The fitting parameters for $\rho = 19.73$ nm$^{-2}$ are $\alpha_2 = 0.058\pm0.004$~K$^{-2}$, $\beta_1 = 1020\pm440$, $\epsilon_1 = 9.85\pm0.63$~K, $\beta_2 = 5.1\pm1.5$ and $\epsilon_2 = 3.69\pm0.21$~K.
It is intuitive to note that $\epsilon_1/\epsilon_0 = 1.6\pm0.1$ is fairly close to the measured enhancement of the peak temperature, $T_{\mathrm{peak}}$($^4$He)/$T_{\mathrm{peak}}$($^3$He)~$= 1.41\pm0.01$.
This can naturally be understood within the theory by Gopalakrishnan, Teo and Hughes~\cite{Gopalakrishnan2013PRL_SM} as a result of intertwining between the hexatic and superfluid orders in a QLC with a triangular lattice.
According to their theory, $T_{\mathrm{i}}$ is determined by unbinding of the $\pm\pi/3$ disclination dipole.
For spatial and phase configurations around this defect, see Fig.~4(e) of Ref.~[\citenum{Gopalakrishnan2013PRL_SM}].
It is important to notice that both $\epsilon_0$ of $^3$He and $\epsilon_1$ of $^4$He are much larger than the potential corrugation ($\leq$ 3 K) indicating that the activation energies are not determined by the corrugation.

We remark that Eq.~(\ref{eq4}) also reproduces the $^4$He data as well as Eq.~(\ref{eq3}).
A fitted $n$ value averaged over densities within the pure $^4$He-C2 phase is 4.90$\pm$0.04 which is close to 5.
It might be related to the anomalously large $T^7$ contribution to the specific heat measured in hcp $^4$He~\cite{Gardner1973PRA}.
This is because the anomaly is interpreted as a signature of the ZP defectons (ZPDs)~\cite{Anderson2005Science} or anomalous phonon dispersion~\cite{Maris2007JLTP}, and the former theory predicts a large $T^5$ contribution in 2D.
On the other hand, fitting of the $^3$He data to Eq.~(\ref{eq4}) gives $n = 8.0\pm$0.2 which is far from the expected value ($= 5$).

It is not clear if the low-$T$ specific heat of $^4$He obeys the $T^2$ law because of the too small heat capacity compared to the addendum.
What we can safely claim is the fact that the $^4$He-C2 phase has significantly (a factor of 3) smaller specific heat than $^3$He at 0.3 $\leq T\leq$ 0.6~K as tabulated in TABLE~\ref{tabI} and shown in the inset of Fig.3(b) of the main text.
Also, $\alpha_2$ does not change appreciably throughout the density region for the pure C2-phase (19.6 $\leq \rho \leq$ 20.3~nm$^{-2}$).
This is again an anomalous isotope effect if we notice that in other 2D and 3D solids of He~\cite{Hering1976JLTP,SampleSwenson1967PR,EdwardsPandorf1965PR} heavier $^4$He have always larger $\alpha_2$ or $\alpha_3$ than lighter $^3$He with the same densities by 40\%--60\% (see TABLE~\ref{tabI}).
Here $\alpha_3$ is the coefficient of the phonon $T^3$ term in the low-$T$ specific heat of 3D solid.
It is likely that the number of phonon modes in the $^4$He-C2 phase is smaller than that of $^3$He being supportive of the QLC picture.
Superfluidity would also reduce $\alpha_2$ significantly as in the superfluid phase of liq. $^4$He in 3D due to coherent motion of bosonic particles.

\renewcommand{\thetable}{S\Roman{table}}
\begin{table*}[t]
\begin{center}
\caption{Isotope effects on melting and phonon contribution to the specific heat in 2D and 3D solid He. C1 and IC1 are the commensurate and incommensurate solids in the first layer of He adsorbed on graphite, respectively. C2 is the second-layer phase which is of the main interest in the present work. IC2 is the incommensurate solid in the second layer. $\rho_1$, $\rho_2$, $T_{\mathrm{peak}}$ and $\alpha_2$ are the first-layer areal density, second-layer density, specific-heat peak temperature and coefficient of the $T^2$ term in the low-$T$ specific heat, respectively, in 2D solid. $V$,  $T_{\mathrm{m}}$ and $\alpha_3$ are the molar volume, melting temperature and coefficient  of the $T^3$ term in the low-$T$ specific heat, respectively, in bulk solid He. The C2 phase has anomalous isotope effects on $T_{\mathrm{peak}}$ and $\alpha_2$ with opposite signs to those of other quantum solids.   
\label{tabI}}

\begin{tabular}{lllllllll} \\
\hline \hline
phase & isotope & ~$\rho_1$ or $\rho_2$& ~$T_{\mathrm{peak}}$ & ~~~$\alpha_2$ & ~~~~~~~~$V$ & ~~$T_{\mathrm{m}}$ & ~~~$\alpha_3$ & ~~~reference \\
 &  & (nm$^{-2}$) & ~~(K) & ~(K$^{-2}$) & ~~~(cm$^3/$mol) & ~~(K) & ~(K$^{-3}$) &  \\
 \hline
C1 & $^3$He & ~~6.37 & 3.02(1) & ~~------ & ~~~~~~------ & ~------ & ~~------ & ~~~Ref.~[\citenum{Bretz1977PRL}] \\
C1 & $^4$He	& ~~6.37 & 2.90(1) & ~~------ & ~~~~~~------ & ~------ & ~~------ & ~~~Ref.~[\citenum{Bretz1977PRL}] \\
\\
IC1 & $^3$He	 & ~~7.8 & 1.26(1) & ~~0.10(2) & ~~~~~~------ & ~------ & ~~------ & ~~~Ref.~[\citenum{Hering1976JLTP}] \\
IC1 & $^4$He	 & ~~7.8 & 1.22(2) & ~~0.14(0) & ~~~~~~------ & ~------ & ~~------ & ~~~Refs.~[\citenum{Hering1976JLTP,Bretz1973PRA,Greywall1993PRB}] \\
\\
C2 & $^3$He	& ~~7.4 & 1.042(4) & ~~0.154(4) & ~~~~~~------ & ~------ & ~~------ & ~~~This work \\
C2 & $^3$He	& ~~7.7 & 1.118(5) & ~~0.249(9) & ~~~~~~------ & ~------ & ~~------ & ~~~This work \\
C2 & $^4$He	& ~~7.6--8.2 & 1.44--1.55 & ~~0.055(11)\footnotemark[1] & ~~~~~~------ & ~------ & ~~------ & ~~~This work \\
\\
IC2 & $^3$He	 & ~~8.5\footnotemark[2] &1.01(1) & ~~------ & ~~~~~~------ & ~------ & ~~------ & ~~~Ref.~[\citenum{VanSciver1978PRB_SM}] \\
IC2 & $^4$He	 & ~~8.5 & 0.99(4) & ~~------ & ~~~~~~------ & ~------ & ~~------ & ~~~This work \\
\\
bcc & $^3$He & ~~------ & ~------ & ~~------ & ~~~~~~20.9 & 2.06(1) & ~~------ & ~~~Ref.~[\citenum{Grilly1959AP}] \\
bcc & $^4$He & ~~------ & ~------ & ~~------ & ~~~~~~20.9 & 1.64(3) & ~~------ & ~~~Ref.~[\citenum{Grilly1959AP,Grilly1973JLTP}] \\
\\
hcp & $^3$He	 & ~~------ & ~------ & ~~------ & ~~~~~~19.2 & 2.98(1) & ~~0.0042(2) & ~~~Ref.~[\citenum{Grilly1959AP,SampleSwenson1967PR}] \\
hcp & $^4$He & ~~------ & ~------ & ~~------ & ~~~~~~19.2 & 2.59(1) & ~~0.0066(2) & ~~~Ref.~[\citenum{Grilly1959AP,EdwardsPandorf1965PR}] \\
\hline \hline
\multicolumn{9}{l}{\footnotesize{$^{\text{a}}$A clear $T^2$ behavior was not seen in the $^4$He-C2 phase.}}\\
\multicolumn{9}{l}{\footnotesize{$^{\text{b}}$$\rho_2$ for the $^3$He-IC2 phase was estimated from the $\rho_2$ vs. $\rho$ relation proposed in Ref.\citenum{Roger1998JLTP}.}}
\end{tabular} 
\end{center}
\end{table*}

\renewcommand{\thefigure}{S\arabic{figure}}
\begin{figure}[t]
\begin{center}
\includegraphics[width=.96\textwidth]{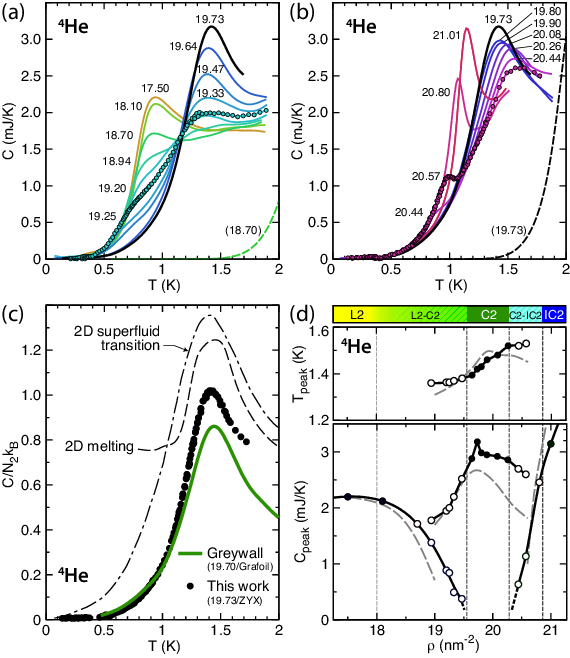}%
\caption{Larger representation of Fig. 1 in the main text.
\label{fig1_SM}}
\end{center}
\end{figure}

\begin{figure}[t]
\begin{center}
\includegraphics[width=.96\textwidth]{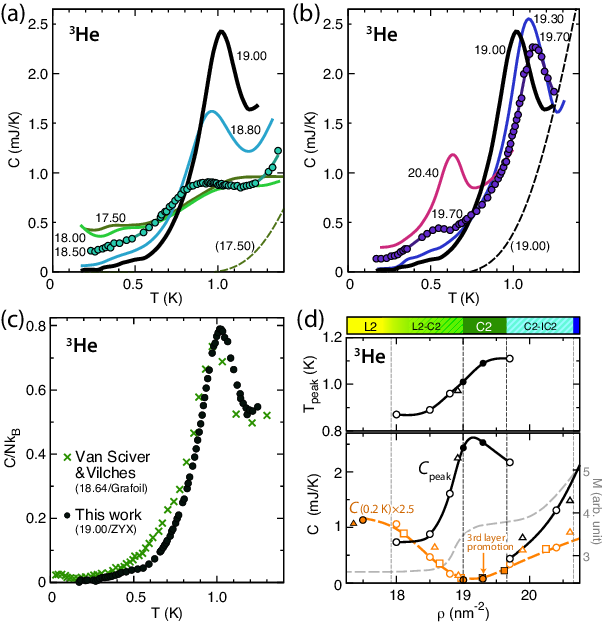}%
\caption{Larger representation of Fig. 2 in the main text.
\label{fig2_SM}}
\end{center}
\end{figure}

\begin{figure}[t]
\begin{center}
\includegraphics[width=.96\textwidth]{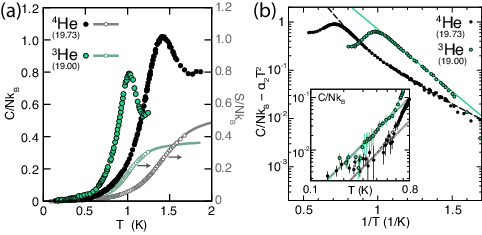}%
\caption{Larger representation of Fig. 3 in the main text.
\label{fig3_SM}}
\end{center}
\end{figure}



\begin{thebibliography}{58}%
\makeatletter
\providecommand \@ifxundefined [1]{%
 \@ifx{#1\undefined}
}%
\providecommand \@ifnum [1]{%
 \ifnum #1\expandafter \@firstoftwo
 \else \expandafter \@secondoftwo
 \fi
}%
\providecommand \@ifx [1]{%
 \ifx #1\expandafter \@firstoftwo
 \else \expandafter \@secondoftwo
 \fi
}%
\providecommand \natexlab [1]{#1}%
\providecommand \enquote  [1]{``#1''}%
\providecommand \bibnamefont  [1]{#1}%
\providecommand \bibfnamefont [1]{#1}%
\providecommand \citenamefont [1]{#1}%
\providecommand \href@noop [0]{\@secondoftwo}%
\providecommand \href [0]{\begingroup \@sanitize@url \@href}%
\providecommand \@href[1]{\@@startlink{#1}\@@href}%
\providecommand \@@href[1]{\endgroup#1\@@endlink}%
\providecommand \@sanitize@url [0]{\catcode `\\12\catcode `\$12\catcode
  `\&12\catcode `\#12\catcode `\^12\catcode `\_12\catcode `\%12\relax}%
\providecommand \@@startlink[1]{}%
\providecommand \@@endlink[0]{}%
\providecommand \url  [0]{\begingroup\@sanitize@url \@url }%
\providecommand \@url [1]{\endgroup\@href {#1}{\urlprefix }}%
\providecommand \urlprefix  [0]{URL }%
\providecommand \Eprint [0]{\href }%
\providecommand \doibase [0]{http://dx.doi.org/}%
\providecommand \selectlanguage [0]{\@gobble}%
\providecommand \bibinfo  [0]{\@secondoftwo}%
\providecommand \bibfield  [0]{\@secondoftwo}%
\providecommand \translation [1]{[#1]}%
\providecommand \BibitemOpen [0]{}%
\providecommand \bibitemStop [0]{}%
\providecommand \bibitemNoStop [0]{.\EOS\space}%
\providecommand \EOS [0]{\spacefactor3000\relax}%
\providecommand \BibitemShut  [1]{\csname bibitem#1\endcsname}%
\let\auto@bib@innerbib\@empty
\bibitem [{\citenamefont {Fradkin}\ \emph {et~al.}(2010)\citenamefont
  {Fradkin}, \citenamefont {Kivelson}, \citenamefont {Lawler}, \citenamefont
  {Eisenstein},\ and\ \citenamefont {Mackenzie}}]{Fradkin2010ARCMP}%
  \BibitemOpen
  \bibfield  {author} {\bibinfo {author} {\bibfnamefont {E.}~\bibnamefont
  {Fradkin}}, \bibinfo {author} {\bibfnamefont {S.~A.}\ \bibnamefont
  {Kivelson}}, \bibinfo {author} {\bibfnamefont {M.~J.}\ \bibnamefont
  {Lawler}}, \bibinfo {author} {\bibfnamefont {J.~P.}\ \bibnamefont
  {Eisenstein}}, \ and\ \bibinfo {author} {\bibfnamefont {A.~P.}\ \bibnamefont
  {Mackenzie}},\ }\href@noop {} {\bibfield  {journal} {\bibinfo  {journal}
  {Annu. Rev. Condens. Matter Phys.}\ }\textbf {\bibinfo {volume} {1}},\ \bibinfo
  {pages} {153} (\bibinfo {year} {2010})}\BibitemShut {NoStop}%
\bibitem [{\citenamefont {Frey}\ \emph {et~al.}(1994)\citenamefont {Frey},
  \citenamefont {Nelson},\ and\ \citenamefont {Fisher}}]{Frey1994PRB}%
  \BibitemOpen
  \bibfield  {author} {\bibinfo {author} {\bibfnamefont {E.}~\bibnamefont
  {Frey}}, \bibinfo {author} {\bibfnamefont {D.~R.}\ \bibnamefont {Nelson}}, \
  and\ \bibinfo {author} {\bibfnamefont {D.~S.}\ \bibnamefont {Fisher}},\
  }\href {\doibase 10.1103/PhysRevB.49.9723} {\bibfield  {journal} {\bibinfo
  {journal} {Phys. Rev. B}\ }\textbf {\bibinfo {volume} {49}},\ \bibinfo
  {pages} {9723} (\bibinfo {year} {1994})}\BibitemShut {NoStop}%
\bibitem [{\citenamefont {Andreev}\ and\ \citenamefont
  {Lifshitz}(1969)}]{Andreev1969JETP}%
  \BibitemOpen
  \bibfield  {author} {\bibinfo {author} {\bibfnamefont {A.~F.}\ \bibnamefont
  {Andreev}}\ and\ \bibinfo {author} {\bibfnamefont {I.~M.}\ \bibnamefont
  {Lifshitz}},\ }\href@noop {} {\bibfield  {journal} {\bibinfo  {journal} {Sov.
  Phys. JETP}\ }\textbf {\bibinfo {volume} {29}},\ \bibinfo {pages} {1107}
  (\bibinfo {year} {1969})}\BibitemShut {NoStop}%
\bibitem [{\citenamefont {Kim}\ and\ \citenamefont
  {Chan}(2004)}]{Kim2004Nature}%
  \BibitemOpen
  \bibfield  {author} {\bibinfo {author} {\bibfnamefont {E.}~\bibnamefont
  {Kim}}\ and\ \bibinfo {author} {\bibfnamefont {M.~H.~W.}\ \bibnamefont
  {Chan}},\ }\href {\doibase 10.1038/nature02220} {\bibfield  {journal}
  {\bibinfo  {journal} {Nature (London)}\ }\textbf {\bibinfo {volume} {427}},\ \bibinfo
  {pages} {225} (\bibinfo {year} {2004})}\BibitemShut {NoStop}%
\bibitem [{\citenamefont {Kim}\ and\ \citenamefont {Chan}(2012)}]{Kim2012PRL}%
  \BibitemOpen
  \bibfield  {author} {\bibinfo {author} {\bibfnamefont {D.~Y.}\ \bibnamefont
  {Kim}}\ and\ \bibinfo {author} {\bibfnamefont {M.~H.~W.}\ \bibnamefont
  {Chan}},\ }\href {\doibase 10.1103/PhysRevLett.109.155301} {\bibfield
  {journal} {\bibinfo  {journal} {Phys. Rev. Lett.}\ }\textbf {\bibinfo
  {volume} {109}},\ \bibinfo {pages} {155301} (\bibinfo {year}
  {2012})}\BibitemShut {NoStop}%
\bibitem [{\citenamefont {Balibar}\ and\ \citenamefont
  {Caupin}(2008)}]{Balibar2008JP}%
  \BibitemOpen
  \bibfield  {author} {\bibinfo {author} {\bibfnamefont {S.}~\bibnamefont
  {Balibar}}\ and\ \bibinfo {author} {\bibfnamefont {F.}~\bibnamefont
  {Caupin}},\ }\href {http://stacks.iop.org/0953-8984/20/i=17/a=173201}
  {\bibfield  {journal} {\bibinfo  {journal} {J. Phys.: Condens. Matter}\
  }\textbf {\bibinfo {volume} {20}},\ \bibinfo {pages} {173201} (\bibinfo
  {year} {2008})}\BibitemShut {NoStop}%
\bibitem [{\citenamefont {Vekhov}\ \emph {et~al.}(2014)\citenamefont {Vekhov},
  \citenamefont {Mullin},\ and\ \citenamefont {Hallock}}]{Vekhov2014PRL}%
  \BibitemOpen
  \bibfield  {author} {\bibinfo {author} {\bibfnamefont {Y.}~\bibnamefont
  {Vekhov}}, \bibinfo {author} {\bibfnamefont {W.~J.}\ \bibnamefont {Mullin}},
  \ and\ \bibinfo {author} {\bibfnamefont {R.~B.}\ \bibnamefont {Hallock}},\
  }\href {\doibase 10.1103/PhysRevLett.113.035302} {\bibfield  {journal}
  {\bibinfo  {journal} {Phys. Rev. Lett.}\ }\textbf {\bibinfo {volume} {113}},\
  \bibinfo {pages} {035302} (\bibinfo {year} {2014})}\BibitemShut {NoStop}%
\bibitem [{cSS()}]{cSScomment}%
  \BibitemOpen
  \href@noop {} {}\bibinfo {note} {these articles revealed important roles of
  superflow along macroscopic or microscopic defects rather than bulk flow in
  hcp solid $^4$He.}\BibitemShut {Stop}%
\bibitem [{\citenamefont {Wu}\ \emph {et~al.}(2016)\citenamefont {Wu},
  \citenamefont {Block},\ and\ \citenamefont {Bruun}}]{Wu2016SciRep}%
  \BibitemOpen
  \bibfield  {author} {\bibinfo {author} {\bibfnamefont {Z.}~\bibnamefont
  {Wu}}, \bibinfo {author} {\bibfnamefont {J.~K.}\ \bibnamefont {Block}}, \
  and\ \bibinfo {author} {\bibfnamefont {G.~M.}\ \bibnamefont {Bruun}},\ }\href
  {http://dx.doi.org/10.1038/srep19038} {\bibfield  {journal} {\bibinfo
  {journal} {Sci. Rep.}\ }\textbf {\bibinfo {volume} {6}},\ \bibinfo {pages}
  {19038} (\bibinfo {year} {2016})}\BibitemShut {NoStop}%
\bibitem [{\citenamefont {Vorontsov}\ and\ \citenamefont
  {Sauls}(2007)}]{Vorontsov2007PRL}%
  \BibitemOpen
  \bibfield  {author} {\bibinfo {author} {\bibfnamefont {A.~B.}\ \bibnamefont
  {Vorontsov}}\ and\ \bibinfo {author} {\bibfnamefont {J.~A.}\ \bibnamefont
  {Sauls}},\ }\href {\doibase 10.1103/PhysRevLett.98.045301} {\bibfield
  {journal} {\bibinfo  {journal} {Phys. Rev. Lett.}\ }\textbf {\bibinfo
  {volume} {98}},\ \bibinfo {pages} {045301} (\bibinfo {year}
  {2007})}\BibitemShut {NoStop}%
\bibitem [{\citenamefont {Levitin}\ \emph {et~al.}(2013)\citenamefont
  {Levitin}, \citenamefont {Bennett1}, \citenamefont {Casey}, \citenamefont
  {Cowan}, \citenamefont {Saunders}, \citenamefont {Drung}, \citenamefont
  {Schurig},\ and\ \citenamefont {Parpia}}]{Levitin2013Science}%
  \BibitemOpen
  \bibfield  {author} {\bibinfo {author} {\bibfnamefont {L.~V.}\ \bibnamefont
  {Levitin}}, \bibinfo {author} {\bibfnamefont {R.~G.}\ \bibnamefont
  {Bennett}}, \bibinfo {author} {\bibfnamefont {A.}~\bibnamefont {Casey}},
  \bibinfo {author} {\bibfnamefont {B.}~\bibnamefont {Cowan}}, \bibinfo
  {author} {\bibfnamefont {J.}~\bibnamefont {Saunders}}, \bibinfo {author}
  {\bibfnamefont {D.}~\bibnamefont {Drung}}, \bibinfo {author} {\bibfnamefont
  {T.}~\bibnamefont {Schurig}}, \ and\ \bibinfo {author} {\bibfnamefont
  {J.~M.}\ \bibnamefont {Parpia}},\ }\href@noop {} {\bibfield  {journal}
  {\bibinfo  {journal} {Science}\ }\textbf {\bibinfo {volume} {340}},\ \bibinfo
  {pages} {841} (\bibinfo {year} {2013})}\BibitemShut {NoStop}%
\bibitem [{\citenamefont {Ishida}\ \emph {et~al.}(1997)\citenamefont {Ishida},
  \citenamefont {Morishita}, \citenamefont {Yawata},\ and\ \citenamefont
  {Fukuyama}}]{Ishida1997PRL}%
  \BibitemOpen
  \bibfield  {author} {\bibinfo {author} {\bibfnamefont {K.}~\bibnamefont
  {Ishida}}, \bibinfo {author} {\bibfnamefont {M.}~\bibnamefont {Morishita}},
  \bibinfo {author} {\bibfnamefont {K.}~\bibnamefont {Yawata}}, \ and\ \bibinfo
  {author} {\bibfnamefont {H.}~\bibnamefont {Fukuyama}},\ }\href {\doibase
  10.1103/PhysRevLett.79.3451} {\bibfield  {journal} {\bibinfo  {journal}
  {Phys. Rev. Lett.}\ }\textbf {\bibinfo {volume} {79}},\ \bibinfo {pages}
  {3451} (\bibinfo {year} {1997})}\BibitemShut {NoStop}%
\bibitem [{\citenamefont {Fukuyama}\ and\ \citenamefont
  {Morishita}(2000)}]{Fukuyama2000PhysicaB}%
  \BibitemOpen
  \bibfield  {author} {\bibinfo {author} {\bibfnamefont {H.}~\bibnamefont
  {Fukuyama}}\ and\ \bibinfo {author} {\bibfnamefont {M.}~\bibnamefont
  {Morishita}},\ }\href {\doibase
  http://dx.doi.org/10.1016/S0921-4526(99)01502-1} {\bibfield  {journal}
  {\bibinfo  {journal} {Physica B}\ }\textbf {\bibinfo {volume} {280}},\
  \bibinfo {pages} {104 } (\bibinfo {year} {2000})}\BibitemShut {NoStop}%
\bibitem [{\citenamefont {Fukuyama}(2008)}]{Fukuyama2008JPSJ}%
  \BibitemOpen
  \bibfield  {author} {\bibinfo {author} {\bibfnamefont {H.}~\bibnamefont
  {Fukuyama}},\ }\href {\doibase 10.1143/JPSJ.77.111013} {\bibfield  {journal}
  {\bibinfo  {journal} {J. Phys. Soc. Jpn.}\ }\textbf {\bibinfo {volume}
  {77}},\ \bibinfo {pages} {111013} (\bibinfo {year} {2008})}\BibitemShut
  {NoStop}%
\bibitem [{\citenamefont {Matsumoto}\ \emph {et~al.}(2005)\citenamefont
  {Matsumoto}, \citenamefont {Tsuji}, \citenamefont {Murakawa}, \citenamefont
  {Akisato}, \citenamefont {Kambara},\ and\ \citenamefont
  {Fukuyama}}]{Matsumoto2005JLTP}%
  \BibitemOpen
  \bibfield  {author} {\bibinfo {author} {\bibfnamefont {Y.}~\bibnamefont
  {Matsumoto}}, \bibinfo {author} {\bibfnamefont {D.}~\bibnamefont {Tsuji}},
  \bibinfo {author} {\bibfnamefont {S.}~\bibnamefont {Murakawa}}, \bibinfo
  {author} {\bibfnamefont {H.}~\bibnamefont {Akisato}}, \bibinfo {author}
  {\bibfnamefont {H.}~\bibnamefont {Kambara}}, \ and\ \bibinfo {author}
  {\bibfnamefont {H.}~\bibnamefont {Fukuyama}},\ }\href {\doibase
  10.1007/s10909-005-1562-2} {\bibfield  {journal} {\bibinfo  {journal} {J. Low
  Temp. Phys.}\ }\textbf {\bibinfo {volume} {138}},\ \bibinfo {pages} {271}
  (\bibinfo {year} {2005})}\BibitemShut {NoStop}%
\bibitem [{\citenamefont {Murakawa}\ \emph {et~al.}(2006)\citenamefont
  {Murakawa}, \citenamefont {Akisato}, \citenamefont {Matsumoto}, \citenamefont
  {Tsuji}, \citenamefont {Mukai}, \citenamefont {Kambara},\ and\ \citenamefont
  {Fukuyama}}]{Murakawa2006AIP}%
  \BibitemOpen
  \bibfield  {author} {\bibinfo {author} {\bibfnamefont {S.}~\bibnamefont
  {Murakawa}}, \bibinfo {author} {\bibfnamefont {H.}~\bibnamefont {Akisato}},
  \bibinfo {author} {\bibfnamefont {Y.}~\bibnamefont {Matsumoto}}, \bibinfo
  {author} {\bibfnamefont {D.}~\bibnamefont {Tsuji}}, \bibinfo {author}
  {\bibfnamefont {K.}~\bibnamefont {Mukai}}, \bibinfo {author} {\bibfnamefont
  {H.}~\bibnamefont {Kambara}}, \ and\ \bibinfo {author} {\bibfnamefont
  {H.}~\bibnamefont {Fukuyama}},\ \href@noop{} {} in \textit{Low Temperature 
  Physics:~24th International Conference on Low Temperature Physics}, \ edited by 
  Y. Takano, S. P. Hershfield, S. O. Hill, P. J. Hirschfeld, and A. M. Goldman,\ \bibfield  {journal}
  {\bibinfo  {journal} {AIP Conf. Proc.}\ }  No. {\bibinfo {volume} {850}}
   (AIP, Melville, NY, \bibinfo {year} {2006}), p. \bibinfo {pages} {311}}\BibitemShut {NoStop}%
\bibitem [{\citenamefont {Roger}\ \emph {et~al.}(1998)\citenamefont {Roger},
  \citenamefont {B\"auerle}, \citenamefont {Bunkov}, \citenamefont {Chen},\
  and\ \citenamefont {Godfrin}}]{RogerPRL1998}%
  \BibitemOpen
  \bibfield  {author} {\bibinfo {author} {\bibfnamefont {M.}~\bibnamefont
  {Roger}}, \bibinfo {author} {\bibfnamefont {C.}~\bibnamefont {B\"auerle}},
  \bibinfo {author} {\bibfnamefont {Y.~M.}\ \bibnamefont {Bunkov}}, \bibinfo
  {author} {\bibfnamefont {A.-S.}\ \bibnamefont {Chen}}, \ and\ \bibinfo
  {author} {\bibfnamefont {H.}~\bibnamefont {Godfrin}},\ }\href {\doibase
  10.1103/PhysRevLett.80.1308} {\bibfield  {journal} {\bibinfo  {journal}
  {Phys. Rev. Lett.}\ }\textbf {\bibinfo {volume} {80}},\ \bibinfo {pages}
  {1308} (\bibinfo {year} {1998})}\BibitemShut {NoStop}%
\bibitem [{\citenamefont {Fuseya}\ and\ \citenamefont
  {Ogata}(2009)}]{Fuseya2009JPSJ}%
  \BibitemOpen
  \bibfield  {author} {\bibinfo {author} {\bibfnamefont {Y.}~\bibnamefont
  {Fuseya}}\ and\ \bibinfo {author} {\bibfnamefont {M.}~\bibnamefont {Ogata}},\
  }\href {\doibase 10.1143/JPSJ.78.013601} {\bibfield  {journal} {\bibinfo
  {journal} {J. Phys. Soc. Jpn.}\ }\textbf {\bibinfo {volume} {78}},\ \bibinfo
  {pages} {013601} (\bibinfo {year} {2009})}\BibitemShut {NoStop}%
\bibitem [{\citenamefont {Watanabe}\ and\ \citenamefont
  {Imada}(2009)}]{Watanabe2009JPSJ}%
  \BibitemOpen
  \bibfield  {author} {\bibinfo {author} {\bibfnamefont {S.}~\bibnamefont
  {Watanabe}}\ and\ \bibinfo {author} {\bibfnamefont {M.}~\bibnamefont
  {Imada}},\ }\href {\doibase 10.1143/JPSJ.78.033603} {\bibfield  {journal}
  {\bibinfo  {journal} {J. Phys. Soc. Jpn.}\ }\textbf {\bibinfo {volume}
  {78}},\ \bibinfo {pages} {033603} (\bibinfo {year} {2009})}\BibitemShut
  {NoStop}%
\bibitem [{\citenamefont {Crowell}\ and\ \citenamefont
  {Reppy}(1996)}]{Crowell1996PRB}%
  \BibitemOpen
  \bibfield  {author} {\bibinfo {author} {\bibfnamefont {P.~A.}\ \bibnamefont
  {Crowell}}\ and\ \bibinfo {author} {\bibfnamefont {J.~D.}\ \bibnamefont
  {Reppy}},\ }\href {\doibase 10.1103/PhysRevB.53.2701} {\bibfield  {journal}
  {\bibinfo  {journal} {Phys. Rev. B}\ }\textbf {\bibinfo {volume} {53}},\
  \bibinfo {pages} {2701} (\bibinfo {year} {1996})}\BibitemShut {NoStop}%
\bibitem [{\citenamefont {Shibayama}\ \emph {et~al.}(2009)\citenamefont
  {Shibayama}, \citenamefont {Fukuyama},\ and\ \citenamefont
  {Shirahama}}]{Shibayama2009JPhys}%
  \BibitemOpen
  \bibfield  {author} {\bibinfo {author} {\bibfnamefont {Y.}~\bibnamefont
  {Shibayama}}, \bibinfo {author} {\bibfnamefont {H.}~\bibnamefont {Fukuyama}},
  \ and\ \bibinfo {author} {\bibfnamefont {K.}~\bibnamefont {Shirahama}},\
  }\href {http://stacks.iop.org/1742-6596/150/i=3/a=032096} {\bibfield
  {journal} {\bibinfo  {journal} {J. Phys.: Conf. Ser.}\ }\textbf {\bibinfo
  {volume} {150}},\ \bibinfo {pages} {032096} (\bibinfo {year}
  {2009})}\BibitemShut {NoStop}%
\bibitem [{\citenamefont {Sato}\ \emph {et~al.}(2012)\citenamefont {Sato},
  \citenamefont {Naruse}, \citenamefont {Matsui},\ and\ \citenamefont
  {Fukuyama}}]{Sato2012PRL}%
  \BibitemOpen
  \bibfield  {author} {\bibinfo {author} {\bibfnamefont {D.}~\bibnamefont
  {Sato}}, \bibinfo {author} {\bibfnamefont {K.}~\bibnamefont {Naruse}},
  \bibinfo {author} {\bibfnamefont {T.}~\bibnamefont {Matsui}}, \ and\ \bibinfo
  {author} {\bibfnamefont {H.}~\bibnamefont {Fukuyama}},\ }\href {\doibase
  10.1103/PhysRevLett.109.235306} {\bibfield  {journal} {\bibinfo  {journal}
  {Phys. Rev. Lett.}\ }\textbf {\bibinfo {volume} {109}},\ \bibinfo {pages}
  {235306} (\bibinfo {year} {2012})}\BibitemShut {NoStop}%
\bibitem [{\citenamefont {Greywall}(1993)}]{Greywall1993PRB}%
  \BibitemOpen
  \bibfield  {author} {\bibinfo {author} {\bibfnamefont {D.~S.}\ \bibnamefont
  {Greywall}},\ }\href {\doibase 10.1103/Greywall_PRB1993} {\bibfield
  {journal} {\bibinfo  {journal} {Phys. Rev. B}\ }\textbf {\bibinfo {volume}
  {47}},\ \bibinfo {pages} {309} (\bibinfo {year} {1993})}\BibitemShut
  {NoStop}%
\bibitem [{\citenamefont {Van~Sciver}\ and\ \citenamefont
  {Vilches}(1978)}]{VanSciverVilches1978PRB}%
  \BibitemOpen
  \bibfield  {author} {\bibinfo {author} {\bibfnamefont {S.~W.}\ \bibnamefont
  {Van~Sciver}}\ and\ \bibinfo {author} {\bibfnamefont {O.~E.}\ \bibnamefont
  {Vilches}},\ }\href {\doibase 10.1103/PhysRevB.18.285} {\bibfield  {journal}
  {\bibinfo  {journal} {Phys. Rev. B}\ }\textbf {\bibinfo {volume} {18}},\
  \bibinfo {pages} {285} (\bibinfo {year} {1978})}\BibitemShut {NoStop}%
\bibitem [{\citenamefont {Corboz}\ \emph {et~al.}(2008)\citenamefont {Corboz},
  \citenamefont {Boninsegni}, \citenamefont {Pollet},\ and\ \citenamefont
  {Troyer}}]{Corboz2008PRB}%
  \BibitemOpen
  \bibfield  {author} {\bibinfo {author} {\bibfnamefont {P.}~\bibnamefont
  {Corboz}}, \bibinfo {author} {\bibfnamefont {M.}~\bibnamefont {Boninsegni}},
  \bibinfo {author} {\bibfnamefont {L.}~\bibnamefont {Pollet}}, \ and\ \bibinfo
  {author} {\bibfnamefont {M.}~\bibnamefont {Troyer}},\ }\href {\doibase
  10.1103/PhysRevB.78.245414} {\bibfield  {journal} {\bibinfo  {journal} {Phys.
  Rev. B}\ }\textbf {\bibinfo {volume} {78}},\ \bibinfo {pages} {245414}
  (\bibinfo {year} {2008})}\BibitemShut {NoStop}%
\bibitem [{\citenamefont {Abraham}\ \emph {et~al.}(1990)\citenamefont
  {Abraham}, \citenamefont {Broughton}, \citenamefont {Leung},\ and\
  \citenamefont {Elser}}]{Abraham1990EPL}%
  \BibitemOpen
  \bibfield  {author} {\bibinfo {author} {\bibfnamefont {F.~F.}\ \bibnamefont
  {Abraham}}, \bibinfo {author} {\bibfnamefont {J.~Q.}\ \bibnamefont
  {Broughton}}, \bibinfo {author} {\bibfnamefont {P.~W.}\ \bibnamefont
  {Leung}}, \ and\ \bibinfo {author} {\bibfnamefont {V.}~\bibnamefont
  {Elser}},\ }\href {\doibase 10.1209/0295-5075/12/2/003} {\bibfield  {journal}
  {\bibinfo  {journal} {Europhys. Lett.}\ }\textbf {\bibinfo {volume} {12}},\
  \bibinfo {pages} {107} (\bibinfo {year} {1990})}\BibitemShut {NoStop}%
\bibitem [{\citenamefont {Pierce}\ and\ \citenamefont
  {Manousakis}(1999)}]{Manousakis1999PRB}%
  \BibitemOpen
  \bibfield  {author} {\bibinfo {author} {\bibfnamefont {M.}~\bibnamefont
  {Pierce}}\ and\ \bibinfo {author} {\bibfnamefont {E.}~\bibnamefont
  {Manousakis}},\ }\href {\doibase 10.1103/PhysRevB.59.3802} {\bibfield
  {journal} {\bibinfo  {journal} {Phys. Rev. B}\ }\textbf {\bibinfo {volume}
  {59}},\ \bibinfo {pages} {3802} (\bibinfo {year} {1999})}\BibitemShut
  {NoStop}%
\bibitem [{\citenamefont {Takagi}(2009)}]{Takagi2009JPhys}%
  \BibitemOpen
  \bibfield  {author} {\bibinfo {author} {\bibfnamefont {T.}~\bibnamefont
  {Takagi}},\ }\href {http://stacks.iop.org/1742-6596/150/i=3/a=032102}
  {\bibfield  {journal} {\bibinfo  {journal} {J. Phys.: Conf. Ser.}\ }\textbf
  {\bibinfo {volume} {150}},\ \bibinfo {pages} {032102} (\bibinfo {year}
  {2009})}\BibitemShut {NoStop}%
\bibitem [{\citenamefont {Birgeneau}\ \emph {et~al.}(1982)\citenamefont
  {Birgeneau}, \citenamefont {Heiney},\ and\ \citenamefont
  {Pelz}}]{Birgeneau1982PhysicaBC}%
  \BibitemOpen
  \bibfield  {author} {\bibinfo {author} {\bibfnamefont {R.~J.}\ \bibnamefont
  {Birgeneau}}, \bibinfo {author} {\bibfnamefont {P.~A.}\ \bibnamefont
  {Heiney}}, \ and\ \bibinfo {author} {\bibfnamefont {J.~P.}\ \bibnamefont
  {Pelz}},\ }\href {\doibase http://dx.doi.org/10.1016/0378-4363(82)90540-X}
  {\bibfield  {journal} {\bibinfo  {journal} {Physica B+C}\ }\textbf {\bibinfo
  {volume} {109--110}},\ \bibinfo {pages} {1785} (\bibinfo {year}
  {1982})}\BibitemShut {NoStop}%
\bibitem [{\citenamefont {Niimi}\ \emph {et~al.}(2006)\citenamefont {Niimi},
  \citenamefont {Matsui}, \citenamefont {Kambara}, \citenamefont {Tagami},
  \citenamefont {Tsukada},\ and\ \citenamefont {Fukuyama}}]{Niimi2006PRB}%
  \BibitemOpen
  \bibfield  {author} {\bibinfo {author} {\bibfnamefont {Y.}~\bibnamefont
  {Niimi}}, \bibinfo {author} {\bibfnamefont {T.}~\bibnamefont {Matsui}},
  \bibinfo {author} {\bibfnamefont {H.}~\bibnamefont {Kambara}}, \bibinfo
  {author} {\bibfnamefont {K.}~\bibnamefont {Tagami}}, \bibinfo {author}
  {\bibfnamefont {M.}~\bibnamefont {Tsukada}}, \ and\ \bibinfo {author}
  {\bibfnamefont {H.}~\bibnamefont {Fukuyama}},\ }\href {\doibase
  10.1103/PhysRevB.73.085421} {\bibfield  {journal} {\bibinfo  {journal} {Phys.
  Rev. B}\ }\textbf {\bibinfo {volume} {73}},\ \bibinfo {pages} {085421}
  (\bibinfo {year} {2006})}\BibitemShut {NoStop}%
\bibitem [{\citenamefont {Strandburg}(1988)}]{Strandburg1988RMP}%
  \BibitemOpen
  \bibfield  {author} {\bibinfo {author} {\bibfnamefont {K.~J.}\ \bibnamefont
  {Strandburg}},\ }\href {\doibase 10.1103/RevModPhys.60.161} {\bibfield
  {journal} {\bibinfo  {journal} {Rev. Mod. Phys.}\ }\textbf {\bibinfo {volume}
  {60}},\ \bibinfo {pages} {161} (\bibinfo {year} {1988})}\BibitemShut
  {NoStop}%
\bibitem [{\citenamefont {Nakamura}\ \emph {et~al.}(2012)\citenamefont
  {Nakamura}, \citenamefont {Matsui}, \citenamefont {Matsui},\ and\
  \citenamefont {Fukuyama}}]{lt26nakamura}%
  \BibitemOpen
  \bibfield  {author} {\bibinfo {author} {\bibfnamefont {S.}~\bibnamefont
  {Nakamura}}, \bibinfo {author} {\bibfnamefont {K.}~\bibnamefont {Matsui}},
  \bibinfo {author} {\bibfnamefont {T.}~\bibnamefont {Matsui}}, \ and\ \bibinfo
  {author} {\bibfnamefont {H.}~\bibnamefont {Fukuyama}},\ }\href {\doibase
  10.1088/1742-6596/400/3/032061} {\bibfield  {journal} {\bibinfo  {journal}
  {J. Phys.: Conf. Ser.}\ }\textbf {\bibinfo {volume} {400}},\ \bibinfo {pages}
  {032061} (\bibinfo {year} {2012})}\BibitemShut {NoStop}%
\bibitem [{\citenamefont {Nakamura}\ \emph {et~al.}(2013)\citenamefont
  {Nakamura}, \citenamefont {Matsui}, \citenamefont {Matsui},\ and\
  \citenamefont {Fukuyama}}]{qfs2012nakamura}%
  \BibitemOpen
  \bibfield  {author} {\bibinfo {author} {\bibfnamefont {S.}~\bibnamefont
  {Nakamura}}, \bibinfo {author} {\bibfnamefont {K.}~\bibnamefont {Matsui}},
  \bibinfo {author} {\bibfnamefont {T.}~\bibnamefont {Matsui}}, \ and\ \bibinfo
  {author} {\bibfnamefont {H.}~\bibnamefont {Fukuyama}},\ }\href {\doibase
  10.1007/s10909-012-0847-5} {\bibfield  {journal} {\bibinfo  {journal} {J. Low
  Temp. Phys.}\ }\textbf {\bibinfo {volume} {171}},\ \bibinfo {pages} {711}
  (\bibinfo {year} {2013})}\BibitemShut {NoStop}%
\bibitem [{dEx()}]{dExperimentalComment}%
  \BibitemOpen
  \href@noop {} {}\bibinfo {note} {the $T$-dependent correction applied to the
  raw heat-capacity data above 2.2 K in this reference is unnecessary in the
  present experiment.}\BibitemShut {Stop}%
\bibitem [{\citenamefont {Nakamura}\ \emph {et~al.}()\citenamefont {Nakamura},
  \citenamefont {Matsui}, \citenamefont {Matsui},\ and\ \citenamefont
  {Fukuyama}}]{Nakamura}%
  \BibitemOpen
  \bibfield  {author} {\bibinfo {author} {\bibfnamefont {S.}~\bibnamefont
  {Nakamura}}, \bibinfo {author} {\bibfnamefont {K.}~\bibnamefont {Matsui}},
  \bibinfo {author} {\bibfnamefont {T.}~\bibnamefont {Matsui}}, \ and\ \bibinfo
  {author} {\bibfnamefont {H.}~\bibnamefont {Fukuyama}},\ }\href@noop {}
  {\bibinfo  {journal} {(unpublished)}\ }\BibitemShut {NoStop}%
\bibitem [{\citenamefont {Lauter}\ \emph {et~al.}(1987)\citenamefont {Lauter},
  \citenamefont {Schildberg}, \citenamefont {Godfrin}, \citenamefont
  {Wiechert},\ and\ \citenamefont {Haensel}}]{Lauter1987CanJP}%
  \BibitemOpen
\bibfield  {journal} {  }\bibfield  {author} {\bibinfo {author} {\bibfnamefont
  {H.~J.}\ \bibnamefont {Lauter}}, \bibinfo {author} {\bibfnamefont {H.~P.}\
  \bibnamefont {Schildberg}}, \bibinfo {author} {\bibfnamefont
  {H.}~\bibnamefont {Godfrin}}, \bibinfo {author} {\bibfnamefont
  {H.}~\bibnamefont {Wiechert}}, \ and\ \bibinfo {author} {\bibfnamefont
  {R.}~\bibnamefont {Haensel}},\ }\href@noop {} {\bibfield  {journal} {\bibinfo
   {journal} {Can. J. Phys.}\ }\textbf {\bibinfo {volume} {65}},\ \bibinfo
  {pages} {1435} (\bibinfo {year} {1987})}\BibitemShut {NoStop}%
\bibitem [{\citenamefont {Lauter}\ \emph {et~al.}(1991)\citenamefont {Lauter},
  \citenamefont {Godfrin}, \citenamefont {Frank},\ and\ \citenamefont
  {Leiderer}}]{Lauter1991PTSF}%
  \BibitemOpen
  \bibfield  {author} {\bibinfo {author} {\bibfnamefont {H.~J.}\ \bibnamefont
  {Lauter}}, \bibinfo {author} {\bibfnamefont {H.}~\bibnamefont {Godfrin}},
  \bibinfo {author} {\bibfnamefont {V.~L.~P.}\ \bibnamefont {Frank}}, \ and\
  \bibinfo {author} {\bibfnamefont {P.}~\bibnamefont {Leiderer}},\ }\href@noop
  {} {} in \textit{Phase Transitions
  in Surface Films 2}, edited by\ \bibinfo {editor} {\bibfnamefont {H.}~\bibnamefont {Taub}},
  \bibinfo {editor} {\bibfnamefont {G.}~\bibnamefont {Torzo}}, \bibinfo
  {editor} {\bibfnamefont {H.~J.}\ \bibnamefont {Lauter}}, \ and\ \bibinfo
  {editor} {\bibfnamefont {J.}~\bibnamefont {S.~C.~Fain}}\\ (\bibinfo  {publisher} {Plenum},\ \bibinfo
  {address} {New York},\ \bibinfo {year} {1991}),\ pp.\ \bibinfo {pages}
  {135--151}\BibitemShut {NoStop}%
\bibitem [{\citenamefont {Wierschem}\ and\ \citenamefont
  {Manousakis}(2011)}]{WierschemManousakis2011PRB}%
  \BibitemOpen
  \bibfield  {author} {\bibinfo {author} {\bibfnamefont {K.}~\bibnamefont
  {Wierschem}}\ and\ \bibinfo {author} {\bibfnamefont {E.}~\bibnamefont
  {Manousakis}},\ }\href {\doibase 10.1103/PhysRevB.83.214108} {\bibfield
  {journal} {\bibinfo  {journal} {Phys. Rev. B}\ }\textbf {\bibinfo {volume}
  {83}},\ \bibinfo {pages} {214108} (\bibinfo {year} {2011})}\BibitemShut
  {NoStop}%
\bibitem [{\citenamefont {Ceperley}\ and\ \citenamefont
  {Pollock}(1989)}]{Ceperley1989PRB}%
  \BibitemOpen
  \bibfield  {author} {\bibinfo {author} {\bibfnamefont {D.~M.}\ \bibnamefont
  {Ceperley}}\ and\ \bibinfo {author} {\bibfnamefont {E.~L.}\ \bibnamefont
  {Pollock}},\ }\href {\doibase 10.1103/PhysRevB.39.2084} {\bibfield  {journal}
  {\bibinfo  {journal} {Phys. Rev. B}\ }\textbf {\bibinfo {volume} {39}},\
  \bibinfo {pages} {2084} (\bibinfo {year} {1989})}\BibitemShut {NoStop}%
\bibitem [{\citenamefont {Van~Sciver}(1978)}]{VanSciver1978PRB}%
  \BibitemOpen
  \bibfield  {author} {\bibinfo {author} {\bibfnamefont {S.~W.}\ \bibnamefont
  {Van~Sciver}},\ }\href {\doibase 10.1103/PhysRevB.18.277} {\bibfield
  {journal} {\bibinfo  {journal} {Phys. Rev. B}\ }\textbf {\bibinfo {volume}
  {18}},\ \bibinfo {pages} {277} (\bibinfo {year} {1978})}\BibitemShut
  {NoStop}%
\bibitem [{\citenamefont {Greywall}(1990)}]{Greywall1990PRB}%
  \BibitemOpen
  \bibfield  {author} {\bibinfo {author} {\bibfnamefont {D.~S.}\ \bibnamefont
  {Greywall}},\ }\href {\doibase 10.1103/PhysRevB.41.1842} {\bibfield
  {journal} {\bibinfo  {journal} {Phys. Rev. B}\ }\textbf {\bibinfo {volume}
  {41}},\ \bibinfo {pages} {1842} (\bibinfo {year} {1990})}\BibitemShut
  {NoStop}%
\bibitem [{eGr()}]{eGreywallComment}%
  \BibitemOpen
  \href@noop {} {}\bibinfo {note} {note that the surface area in this reference
  has been corrected by $-2.5$\%. See D. S. Greywall and P. A. Busch, Phys.
  Rev. Lett. {\bf65}, 2788 (1990).}\BibitemShut {Stop}%
\bibitem [{\citenamefont {B\"auerle}\ \emph {et~al.}(1996)\citenamefont
  {B\"auerle}, \citenamefont {Bunkov}, \citenamefont {Fisher},\ and\
  \citenamefont {Godfrin}}]{Bauerle1996Czech}%
  \BibitemOpen
  \bibfield  {author} {\bibinfo {author} {\bibfnamefont {C.}~\bibnamefont
  {B\"auerle}}, \bibinfo {author} {\bibfnamefont {Y.~M.}\ \bibnamefont
  {Bunkov}}, \bibinfo {author} {\bibfnamefont {S.~N.}\ \bibnamefont {Fisher}},
  \ and\ \bibinfo {author} {\bibfnamefont {H.}~\bibnamefont {Godfrin}},\
  }\href@noop {} {\bibfield  {journal} {\bibinfo  {journal} {Czech. J. Phys.,
  Suppl. S1}\ }\textbf {\bibinfo {volume} {46}},\ \bibinfo {pages} {401}
  (\bibinfo {year} {1996})}\BibitemShut {NoStop}%
\bibitem [{\citenamefont {Halperin}\ and\ \citenamefont
  {Nelson}(1978)}]{Halperin-Nelson1978PRL}%
  \BibitemOpen
  \bibfield  {author} {\bibinfo {author} {\bibfnamefont {B.~I.}\ \bibnamefont
  {Halperin}}\ and\ \bibinfo {author} {\bibfnamefont {D.~R.}\ \bibnamefont
  {Nelson}},\ }\href {\doibase 10.1103/PhysRevLett.41.121} {\bibfield
  {journal} {\bibinfo  {journal} {Phys. Rev. Lett.}\ }\textbf {\bibinfo
  {volume} {41}},\ \bibinfo {pages} {121} (\bibinfo {year} {1978})}\BibitemShut
  {NoStop}%
\bibitem [{\citenamefont {Nelson}\ and\ \citenamefont
  {Halperin}(1979)}]{NelsonHalperin1979PRB}%
  \BibitemOpen
  \bibfield  {author} {\bibinfo {author} {\bibfnamefont {D.~R.}\ \bibnamefont
  {Nelson}}\ and\ \bibinfo {author} {\bibfnamefont {B.~I.}\ \bibnamefont
  {Halperin}},\ }\href {\doibase 10.1103/PhysRevB.19.2457} {\bibfield
  {journal} {\bibinfo  {journal} {Phys. Rev. B}\ }\textbf {\bibinfo {volume}
  {19}},\ \bibinfo {pages} {2457} (\bibinfo {year} {1979})}\BibitemShut
  {NoStop}%
\bibitem [{\citenamefont {Young}(1979)}]{Young1979PRB}%
  \BibitemOpen
  \bibfield  {author} {\bibinfo {author} {\bibfnamefont {A.~P.}\ \bibnamefont
  {Young}},\ }\href {\doibase 10.1103/PhysRevB.19.1855} {\bibfield  {journal}
  {\bibinfo  {journal} {Phys. Rev. B}\ }\textbf {\bibinfo {volume} {19}},\
  \bibinfo {pages} {1855} (\bibinfo {year} {1979})}\BibitemShut {NoStop}%
\bibitem [{jKe()}]{jKentcomment}%
  \BibitemOpen
  \href@noop {} {}\bibinfo {note} {Rather sharp specific heat anomalies [half-width at half maximum (HWHM)
  on low-$T$ side $= 0.06T_{\mathrm{peak}}$] are reported for the KT superfluid
  transition in thin $^4$He films adsorbed on porous media [L. M. Steele
  , C. J. Yeager, and D. Finotello, Phys. Rev. Lett. {\bf71}, 3673 (1993)], while it is much broader
  in the theoretical calculation ($= 0.30T_{\mathrm{peak}}$) in
  Ref.~\citenum{Ceperley1989PRB}. The transition nature could be sensitive to
  the film geometry and surface roughness. The HWHM of our $^4$He\,-C2 anomaly is
  $0.18T_{\mathrm{peak}}$.}\BibitemShut {Stop}%
\bibitem [{\citenamefont {Berker}\ and\ \citenamefont
  {Nelson}(1979)}]{BerkerNelson1979PRB}%
  \BibitemOpen
  \bibfield  {author} {\bibinfo {author} {\bibfnamefont {A.~N.}\ \bibnamefont
  {Berker}}\ and\ \bibinfo {author} {\bibfnamefont {D.~R.}\ \bibnamefont
  {Nelson}},\ }\href {\doibase 10.1103/PhysRevB.19.2488} {\bibfield  {journal}
  {\bibinfo  {journal} {Phys. Rev. B}\ }\textbf {\bibinfo {volume} {19}},\
  \bibinfo {pages} {2488} (\bibinfo {year} {1979})}\BibitemShut {NoStop}%
\bibitem [{bDe()}]{bDensityRatioComment}%
  \BibitemOpen
  \href@noop {} {}\bibinfo {note} {Density ratios between the first and second
  layers are $0.62\pm0.05$ ($^4$He) and $0.65\pm0.03$ ($^3$He) which are closer
  to $2/3 = 0.667$ or $13/19 = 0.684$ rather than $4/7 = 0.571$. Hence our data
  are not consistent with the previous structural assignment of the 4/7
  phase.}\BibitemShut {Stop}%
\bibitem [{\citenamefont {Mullen}\ \emph {et~al.}(1994)\citenamefont {Mullen},
  \citenamefont {Stoof}, \citenamefont {Wallin},\ and\ \citenamefont
  {Girvin}}]{Mullen1994PRL}%
  \BibitemOpen
  \bibfield  {author} {\bibinfo {author} {\bibfnamefont {K.}~\bibnamefont
  {Mullen}}, \bibinfo {author} {\bibfnamefont {H.~T.~C.}\ \bibnamefont
  {Stoof}}, \bibinfo {author} {\bibfnamefont {M.}~\bibnamefont {Wallin}}, \
  and\ \bibinfo {author} {\bibfnamefont {S.~M.}\ \bibnamefont {Girvin}},\
  }\href {\doibase 10.1103/PhysRevLett.72.4013} {\bibfield  {journal} {\bibinfo
   {journal} {Phys. Rev. Lett.}\ }\textbf {\bibinfo {volume} {72}},\ \bibinfo
  {pages} {4013} (\bibinfo {year} {1994})}\BibitemShut {NoStop}%
\bibitem [{\citenamefont {Apaja}\ and\ \citenamefont
  {Saarela}(2008)}]{Apaja2008EPL}%
  \BibitemOpen
  \bibfield  {author} {\bibinfo {author} {\bibfnamefont {V.}~\bibnamefont
  {Apaja}}\ and\ \bibinfo {author} {\bibfnamefont {M.}~\bibnamefont
  {Saarela}},\ }\href {http://stacks.iop.org/0295-5075/84/i=4/a=40003}
  {\bibfield  {journal} {\bibinfo  {journal} {Europhys. Lett.}\ }\textbf
  {\bibinfo {volume} {84}},\ \bibinfo {pages} {40003} (\bibinfo {year}
  {2008})}\BibitemShut {NoStop}%
\bibitem [{fHe()}]{fHexaticComment}%
  \BibitemOpen
  \href@noop {} {}\bibinfo {note} {in this calculation for an ideal 2D $^4$He
  system, the sixfold angular correlation appears at densities above
  6.5~nm$^{-2}$, and the hexatic phase persists metastably up to 7.7~nm$^{-2}$
  although solid is a stable phase above 7.0~nm$^{-2}$. In our experiment, the C2
  correlation appears above $\rho_2 =$ 6.9~nm$^{-2}$ and disappears above
  8.5~nm$^{-2}$ in $^4$He.}\BibitemShut {Stop}%
\bibitem [{\citenamefont {Halinen}\ \emph {et~al.}(2000)\citenamefont
  {Halinen}, \citenamefont {Apaja}, \citenamefont {Gernoth},\ and\
  \citenamefont {Saarela}}]{Halinen2000JLTP}%
  \BibitemOpen
  \bibfield  {author} {\bibinfo {author} {\bibfnamefont {J.}~\bibnamefont
  {Halinen}}, \bibinfo {author} {\bibfnamefont {V.}~\bibnamefont {Apaja}},
  \bibinfo {author} {\bibfnamefont {K.~A.}\ \bibnamefont {Gernoth}}, \ and\
  \bibinfo {author} {\bibfnamefont {M.}~\bibnamefont {Saarela}},\ }\href@noop
  {} {\bibfield  {journal} {\bibinfo  {journal} {J. of Low Temp. Phys.}\
  }\textbf {\bibinfo {volume} {121}},\ \bibinfo {pages} {531} (\bibinfo {year}
  {2000})}\BibitemShut {NoStop}%
\bibitem [{\citenamefont {Krotscheck}\ and\ \citenamefont
  {Lichtenegger}(2015)}]{KrotscheckJLTP2015}%
  \BibitemOpen
  \bibfield  {author} {\bibinfo {author} {\bibfnamefont {E.}~\bibnamefont
  {Krotscheck}}\ and\ \bibinfo {author} {\bibfnamefont {T.}~\bibnamefont
  {Lichtenegger}},\ }\href {\doibase 10.1007/s10909-014-1221-6} {\bibfield
  {journal} {\bibinfo  {journal} {J. of Low Temp. Phys.}\ }\textbf {\bibinfo
  {volume} {178}},\ \bibinfo {pages} {61} (\bibinfo {year} {2015})}\BibitemShut
  {NoStop}%
\bibitem [{\citenamefont {Ahn}\ \emph {et~al.}(2016)\citenamefont {Ahn},
  \citenamefont {Lee},\ and\ \citenamefont {Kwon}}]{Ahn2016PRB}%
  \BibitemOpen
  \bibfield  {author} {\bibinfo {author} {\bibfnamefont {J.}~\bibnamefont
  {Ahn}}, \bibinfo {author} {\bibfnamefont {H.}~\bibnamefont {Lee}}, \ and\
  \bibinfo {author} {\bibfnamefont {Y.}~\bibnamefont {Kwon}},\ }\href {\doibase
  10.1103/PhysRevB.93.064511} {\bibfield  {journal} {\bibinfo  {journal} {Phys.
  Rev. B}\ }\textbf {\bibinfo {volume} {93}},\ \bibinfo {pages} {064511}
  (\bibinfo {year} {2016})}\BibitemShut {NoStop}%
\bibitem [{gSM()}]{gSMComment}%
  \BibitemOpen
  \href@noop {} {}\bibinfo {note} {See Supplemental Material which includes Refs.~[16,26,28,33,41].}\BibitemShut {Stop}%
\bibitem [{\citenamefont {Andreev}(1982)}]{Andreev1982PLTP}%
  \BibitemOpen
  \bibfield  {author} {\bibinfo {author} {\bibfnamefont {A.~F.}\ \bibnamefont
  {Andreev}},\ } in \textit{Progress in Low Temp. Phys.} edited by\ \bibinfo {editor} {\bibfnamefont
  {D.~F.}\ \bibnamefont {Brewer}}  (North-Holland, Amsterdam, 1982),\ Vol.\ \bibinfo 
  {volume} {VIII},\  pp.\ \bibinfo {pages} {67--131}\BibitemShut {NoStop}%
\bibitem [{\citenamefont {Gopalakrishnan}\ \emph {et~al.}(2013)\citenamefont
  {Gopalakrishnan}, \citenamefont {Teo},\ and\ \citenamefont
  {Hughes}}]{Gopalakrishnan2013PRL}%
  \BibitemOpen
  \bibfield  {author} {\bibinfo {author} {\bibfnamefont {S.}~\bibnamefont
  {Gopalakrishnan}}, \bibinfo {author} {\bibfnamefont {J.~C.~Y.}\ \bibnamefont
  {Teo}}, \ and\ \bibinfo {author} {\bibfnamefont {T.~L.}\ \bibnamefont
  {Hughes}},\ }\href {\doibase 10.1103/PhysRevLett.111.025304} {\bibfield
  {journal} {\bibinfo  {journal} {Phys. Rev. Lett.}\ }\textbf {\bibinfo
  {volume} {111}},\ \bibinfo {pages} {025304} (\bibinfo {year}
  {2013})}\BibitemShut {NoStop}%
\end{thebibliography}

\begin{thebibliography}{19}%
\makeatletter
\providecommand \@ifxundefined [1]{%
 \@ifx{#1\undefined}
}%
\providecommand \@ifnum [1]{%
 \ifnum #1\expandafter \@firstoftwo
 \else \expandafter \@secondoftwo
 \fi
}%
\providecommand \@ifx [1]{%
 \ifx #1\expandafter \@firstoftwo
 \else \expandafter \@secondoftwo
 \fi
}%
\providecommand \natexlab [1]{#1}%
\providecommand \enquote  [1]{``#1''}%
\providecommand \bibnamefont  [1]{#1}%
\providecommand \bibfnamefont [1]{#1}%
\providecommand \citenamefont [1]{#1}%
\providecommand \href@noop [0]{\@secondoftwo}%
\providecommand \href [0]{\begingroup \@sanitize@url \@href}%
\providecommand \@href[1]{\@@startlink{#1}\@@href}%
\providecommand \@@href[1]{\endgroup#1\@@endlink}%
\providecommand \@sanitize@url [0]{\catcode `\\12\catcode `\$12\catcode
  `\&12\catcode `\#12\catcode `\^12\catcode `\_12\catcode `\%12\relax}%
\providecommand \@@startlink[1]{}%
\providecommand \@@endlink[0]{}%
\providecommand \url  [0]{\begingroup\@sanitize@url \@url }%
\providecommand \@url [1]{\endgroup\@href {#1}{\urlprefix }}%
\providecommand \urlprefix  [0]{URL }%
\providecommand \Eprint [0]{\href }%
\providecommand \doibase [0]{http://dx.doi.org/}%
\providecommand \selectlanguage [0]{\@gobble}%
\providecommand \bibinfo  [0]{\@secondoftwo}%
\providecommand \bibfield  [0]{\@secondoftwo}%
\providecommand \translation [1]{[#1]}%
\providecommand \BibitemOpen [0]{}%
\providecommand \bibitemStop [0]{}%
\providecommand \bibitemNoStop [0]{.\EOS\space}%
\providecommand \EOS [0]{\spacefactor3000\relax}%
\providecommand \BibitemShut  [1]{\csname bibitem#1\endcsname}%
\let\auto@bib@innerbib\@empty
\bibitem [{\citenamefont {Nagaoka}(1980)}]{Nagaoka1980PTPS}%
  \BibitemOpen
  \bibfield  {author} {\bibinfo {author} {\bibfnamefont {Y.}~\bibnamefont
  {Nagaoka}},\ }\href {\doibase 10.1143/PTPS.69.335} {\bibfield  {journal}
  {\bibinfo  {journal} {Prog. Theor. Phys. Suppl.}\ }\textbf {\bibinfo {volume}
  {69}},\ \bibinfo {pages} {335} (\bibinfo {year} {1980})}\BibitemShut
  {NoStop}%
\bibitem [{\citenamefont {Ceperley}\ \emph {et~al.}(1996)\citenamefont
  {Ceperley}, \citenamefont {Simmons},\ and\ \citenamefont
  {Blasdell}}]{Ceperley1996PRL}%
  \BibitemOpen
  \bibfield  {author} {\bibinfo {author} {\bibfnamefont {D.~M.}\ \bibnamefont
  {Ceperley}}, \bibinfo {author} {\bibfnamefont {R.~O.}\ \bibnamefont
  {Simmons}}, \ and\ \bibinfo {author} {\bibfnamefont {R.~C.}\ \bibnamefont
  {Blasdell}},\ }\href {\doibase 10.1103/PhysRevLett.77.115} {\bibfield
  {journal} {\bibinfo  {journal} {Phys. Rev. Lett.}\ }\textbf {\bibinfo
  {volume} {77}},\ \bibinfo {pages} {115} (\bibinfo {year} {1996})}\BibitemShut
  {NoStop}%
\bibitem [{\citenamefont {Hirashima}\ \emph {et~al.}(2003)\citenamefont
  {Hirashima}, \citenamefont {Momoi},\ and\ \citenamefont
  {Takagi}}]{Hirashima2003JPSJ}%
  \BibitemOpen
  \bibfield  {author} {\bibinfo {author} {\bibfnamefont {D.~S.}\ \bibnamefont
  {Hirashima}}, \bibinfo {author} {\bibfnamefont {T.}~\bibnamefont {Momoi}}, \
  and\ \bibinfo {author} {\bibfnamefont {T.}~\bibnamefont {Takagi}},\ }\href
  {\doibase 10.1143/jpsj.72.1446} {\bibfield  {journal} {\bibinfo  {journal}
  {J. Phys. Soc. Jpn.}\ }\textbf {\bibinfo {volume} {72}},\ \bibinfo {pages}
  {1446} (\bibinfo {year} {2003})}\BibitemShut {NoStop}%
\bibitem [{\citenamefont {Van~Sciver}(1978)}]{VanSciver1978PRB_SM}%
  \BibitemOpen
  \bibfield  {author} {\bibinfo {author} {\bibfnamefont {S.~W.}\ \bibnamefont
  {Van~Sciver}},\ }\href {\doibase 10.1103/PhysRevB.18.277} {\bibfield
  {journal} {\bibinfo  {journal} {Phys. Rev. B}\ }\textbf {\bibinfo {volume}
  {18}},\ \bibinfo {pages} {277} (\bibinfo {year} {1978})}\BibitemShut
  {NoStop}%
\bibitem [{\citenamefont {Wierschem}\ and\ \citenamefont
  {Manousakis}(2011)}]{WierschemManousakis2011PRB_SM}%
  \BibitemOpen
  \bibfield  {author} {\bibinfo {author} {\bibfnamefont {K.}~\bibnamefont
  {Wierschem}}\ and\ \bibinfo {author} {\bibfnamefont {E.}~\bibnamefont
  {Manousakis}},\ }\href {\doibase 10.1103/PhysRevB.83.214108} {\bibfield
  {journal} {\bibinfo  {journal} {Phys. Rev. B}\ }\textbf {\bibinfo {volume}
  {83}},\ \bibinfo {pages} {214108} (\bibinfo {year} {2011})}\BibitemShut
  {NoStop}%
\bibitem [{\citenamefont {Berker}\ and\ \citenamefont
  {Nelson}(1979)}]{BerkerNelson1979PRB_SM}%
  \BibitemOpen
  \bibfield  {author} {\bibinfo {author} {\bibfnamefont {A.~N.}\ \bibnamefont
  {Berker}}\ and\ \bibinfo {author} {\bibfnamefont {D.~R.}\ \bibnamefont
  {Nelson}},\ }\href {\doibase 10.1103/PhysRevB.19.2488} {\bibfield  {journal}
  {\bibinfo  {journal} {Phys. Rev. B}\ }\textbf {\bibinfo {volume} {19}},\
  \bibinfo {pages} {2488} (\bibinfo {year} {1979})}\BibitemShut {NoStop}%
\bibitem [{\citenamefont {Gopalakrishnan}\ \emph {et~al.}(2013)\citenamefont
  {Gopalakrishnan}, \citenamefont {Teo},\ and\ \citenamefont
  {Hughes}}]{Gopalakrishnan2013PRL_SM}%
  \BibitemOpen
  \bibfield  {author} {\bibinfo {author} {\bibfnamefont {S.}~\bibnamefont
  {Gopalakrishnan}}, \bibinfo {author} {\bibfnamefont {J.~C.~Y.}\ \bibnamefont
  {Teo}}, \ and\ \bibinfo {author} {\bibfnamefont {T.~L.}\ \bibnamefont
  {Hughes}},\ }\href {\doibase 10.1103/PhysRevLett.111.025304} {\bibfield
  {journal} {\bibinfo  {journal} {Phys. Rev. Lett.}\ }\textbf {\bibinfo
  {volume} {111}},\ \bibinfo {pages} {025304} (\bibinfo {year}
  {2013})}\BibitemShut {NoStop}%
\bibitem [{\citenamefont {Gardner}\ \emph {et~al.}(1973)\citenamefont
  {Gardner}, \citenamefont {Hoffer},\ and\ \citenamefont
  {Phillips}}]{Gardner1973PRA}%
  \BibitemOpen
  \bibfield  {author} {\bibinfo {author} {\bibfnamefont {W.~R.}\ \bibnamefont
  {Gardner}}, \bibinfo {author} {\bibfnamefont {J.~K.}\ \bibnamefont {Hoffer}},
  \ and\ \bibinfo {author} {\bibfnamefont {N.~E.}\ \bibnamefont {Phillips}},\
  }\href {\doibase 10.1103/PhysRevA.7.1029} {\bibfield  {journal} {\bibinfo
  {journal} {Phys. Rev. A}\ }\textbf {\bibinfo {volume} {7}},\ \bibinfo {pages}
  {1029} (\bibinfo {year} {1973})}\BibitemShut {NoStop}%
\bibitem [{\citenamefont {Anderson}\ \emph {et~al.}(2005)\citenamefont
  {Anderson}, \citenamefont {Brinkman},\ and\ \citenamefont
  {Huse}}]{Anderson2005Science}%
  \BibitemOpen
  \bibfield  {author} {\bibinfo {author} {\bibfnamefont {P.~W.}\ \bibnamefont
  {Anderson}}, \bibinfo {author} {\bibfnamefont {W.~F.}\ \bibnamefont
  {Brinkman}}, \ and\ \bibinfo {author} {\bibfnamefont {D.~A.}\ \bibnamefont
  {Huse}},\ }\href {\doibase 10.1126/science.1118625} {\bibfield  {journal}
  {\bibinfo  {journal} {Science}\ }\textbf {\bibinfo {volume} {310}},\ \bibinfo
  {pages} {1164} (\bibinfo {year} {2005})}\BibitemShut {NoStop}%
\bibitem [{\citenamefont {Maris}\ and\ \citenamefont
  {Balibar}(2007)}]{Maris2007JLTP}%
  \BibitemOpen
  \bibfield  {author} {\bibinfo {author} {\bibfnamefont {H.~J.}\ \bibnamefont
  {Maris}}\ and\ \bibinfo {author} {\bibfnamefont {S.}~\bibnamefont
  {Balibar}},\ }\href {\doibase 10.1007/s10909-007-9347-4} {\bibfield
  {journal} {\bibinfo  {journal} {J. Low Temp. Phys.}\ }\textbf {\bibinfo
  {volume} {147}},\ \bibinfo {pages} {539} (\bibinfo {year}
  {2007})}\BibitemShut {NoStop}%
\bibitem [{\citenamefont {Hering}\ \emph {et~al.}(1976)\citenamefont {Hering},
  \citenamefont {Van~Sciver},\ and\ \citenamefont {Vilches}}]{Hering1976JLTP}%
  \BibitemOpen
  \bibfield  {author} {\bibinfo {author} {\bibfnamefont {S.~V.}\ \bibnamefont
  {Hering}}, \bibinfo {author} {\bibfnamefont {S.~W.}\ \bibnamefont
  {Van~Sciver}}, \ and\ \bibinfo {author} {\bibfnamefont {O.~E.}\ \bibnamefont
  {Vilches}},\ }\href {\doibase 10.1007/BF00657299} {\bibfield  {journal}
  {\bibinfo  {journal} {J. Low Temp. Phys.}\ }\textbf {\bibinfo {volume}
  {25}},\ \bibinfo {pages} {793} (\bibinfo {year} {1976})}\BibitemShut
  {NoStop}%
\bibitem [{\citenamefont {Sample}\ and\ \citenamefont
  {Swenson}(1967)}]{SampleSwenson1967PR}%
  \BibitemOpen
  \bibfield  {author} {\bibinfo {author} {\bibfnamefont {H.~H.}\ \bibnamefont
  {Sample}}\ and\ \bibinfo {author} {\bibfnamefont {C.~A.}\ \bibnamefont
  {Swenson}},\ }\href {\doibase 10.1103/PhysRev.158.188} {\bibfield  {journal}
  {\bibinfo  {journal} {Phys. Rev.}\ }\textbf {\bibinfo {volume} {158}},\
  \bibinfo {pages} {188} (\bibinfo {year} {1967})}\BibitemShut {NoStop}%
\bibitem [{\citenamefont {Edwards}\ and\ \citenamefont
  {Pandorf}(1965)}]{EdwardsPandorf1965PR}%
  \BibitemOpen
  \bibfield  {author} {\bibinfo {author} {\bibfnamefont {D.~O.}\ \bibnamefont
  {Edwards}}\ and\ \bibinfo {author} {\bibfnamefont {R.~C.}\ \bibnamefont
  {Pandorf}},\ }\href {\doibase 10.1103/PhysRev.140.A816} {\bibfield  {journal}
  {\bibinfo  {journal} {Phys. Rev.}\ }\textbf {\bibinfo {volume} {140}},\
  \bibinfo {pages} {A816} (\bibinfo {year} {1965})}\BibitemShut {NoStop}%
\bibitem [{\citenamefont {Bretz}(1977)}]{Bretz1977PRL}%
  \BibitemOpen
  \bibfield  {author} {\bibinfo {author} {\bibfnamefont {M.}~\bibnamefont
  {Bretz}},\ }\href {\doibase 10.1103/Bretz_PRL1977} {\bibfield  {journal}
  {\bibinfo  {journal} {Phys. Rev. Lett.}\ }\textbf {\bibinfo {volume} {38}},\
  \bibinfo {pages} {501} (\bibinfo {year} {1977})}\BibitemShut {NoStop}%
\bibitem [{\citenamefont {Bretz}\ \emph {et~al.}(1973)\citenamefont {Bretz},
  \citenamefont {Dash}, \citenamefont {Hickernell}, \citenamefont {McLean},\
  and\ \citenamefont {Vilches}}]{Bretz1973PRA}%
  \BibitemOpen
  \bibfield  {author} {\bibinfo {author} {\bibfnamefont {M.}~\bibnamefont
  {Bretz}}, \bibinfo {author} {\bibfnamefont {J.~G.}\ \bibnamefont {Dash}},
  \bibinfo {author} {\bibfnamefont {D.~C.}\ \bibnamefont {Hickernell}},
  \bibinfo {author} {\bibfnamefont {E.~O.}\ \bibnamefont {McLean}}, \ and\
  \bibinfo {author} {\bibfnamefont {O.~E.}\ \bibnamefont {Vilches}},\ }\href
  {\doibase 10.1103/PhysRevA.8.1589} {\bibfield  {journal} {\bibinfo  {journal}
  {Phys. Rev. A}\ }\textbf {\bibinfo {volume} {8}},\ \bibinfo {pages} {1589}
  (\bibinfo {year} {1973})}\BibitemShut {NoStop}%
\bibitem [{\citenamefont {Greywall}(1993)}]{Greywall1993PRB}%
  \BibitemOpen
  \bibfield  {author} {\bibinfo {author} {\bibfnamefont {D.~S.}\ \bibnamefont
  {Greywall}},\ }\href {\doibase 10.1103/Greywall_PRB1993} {\bibfield
  {journal} {\bibinfo  {journal} {Phys. Rev. B}\ }\textbf {\bibinfo {volume}
  {47}},\ \bibinfo {pages} {309} (\bibinfo {year} {1993})}\BibitemShut
  {NoStop}%
\bibitem [{\citenamefont {Grilly}\ and\ \citenamefont
  {Mills}(1959)}]{Grilly1959AP}%
  \BibitemOpen
  \bibfield  {author} {\bibinfo {author} {\bibfnamefont {E.~R.}\ \bibnamefont
  {Grilly}}\ and\ \bibinfo {author} {\bibfnamefont {R.~L.}\ \bibnamefont
  {Mills}},\ }\href {\doibase http://dx.doi.org/10.1016/0003-4916(59)90061-2}
  {\bibfield  {journal} {\bibinfo  {journal} {Ann. Phys.}\ }\textbf {\bibinfo
  {volume} {8}},\ \bibinfo {pages} {1 } (\bibinfo {year} {1959})}\BibitemShut
  {NoStop}%
\bibitem [{\citenamefont {Grilly}(1973)}]{Grilly1973JLTP}%
  \BibitemOpen
  \bibfield  {author} {\bibinfo {author} {\bibfnamefont {E.~R.}\ \bibnamefont
  {Grilly}},\ }\href {\doibase 10.1007/BF00655035} {\bibfield  {journal}
  {\bibinfo  {journal} {J. Low Temp. Phys.}\ }\textbf {\bibinfo {volume}
  {11}},\ \bibinfo {pages} {33} (\bibinfo {year} {1973})}\BibitemShut {NoStop}%
\bibitem [{\citenamefont {Roger}\ \emph {et~al.}(1998)\citenamefont {Roger},
  \citenamefont {B\"auerle}, \citenamefont {Godfrin}, \citenamefont
  {Pricoupenko},\ and\ \citenamefont {Treiner}}]{Roger1998JLTP}%
  \BibitemOpen
  \bibfield  {author} {\bibinfo {author} {\bibfnamefont {M.}~\bibnamefont
  {Roger}}, \bibinfo {author} {\bibfnamefont {C.}~\bibnamefont {B\"auerle}},
  \bibinfo {author} {\bibfnamefont {H.}~\bibnamefont {Godfrin}}, \bibinfo
  {author} {\bibfnamefont {L.}~\bibnamefont {Pricoupenko}}, \ and\ \bibinfo
  {author} {\bibfnamefont {J.}~\bibnamefont {Treiner}},\ }\href {\doibase
  10.1023/A:1022344904078} {\bibfield  {journal} {\bibinfo  {journal} {J. Low
  Temp. Phys.}\ }\textbf {\bibinfo {volume} {112}},\ \bibinfo {pages} {451}
  (\bibinfo {year} {1998})}\BibitemShut {NoStop}%
\end{thebibliography}

%

\end{document}